\documentclass[aps,floatfix]{revtex4}
\usepackage{graphicx}
\usepackage{epsfig}
\usepackage{amsmath}
\usepackage{graphics}
\usepackage{latexsym}
\usepackage[usenames]{color}

\begin{document}

\title{Unconventional ordering behavior of semi-flexible polymers in dense
brushes under compression} 

\author{Andrey Milchev$^{1,2}$ and Kurt Binder$^2$}

\affiliation{$^1$ Institute of Physical Chemistry, Bulgarian Academy of
Sciences, Sofia 1113, Bulgaria\\ $^2$Institut f\"ur Physik, Johannes
Gutenberg-Universit\"at Mainz, Staudinger Weg 7, D-55099 Mainz, Germany}

\begin{abstract}
Using a coarse-grained bead-spring model for semi-flexible macromolecules
forming a polymer brush, structure and dynamics of the polymers is investigated,
varying chain stiffness and grafting density. The anchoring condition for the
grafted chains is chosen such that their first bonds are oriented along the
normal to the substrate plane.

Compression of such a semi-flexible brush by a planar piston is observed to be a
two-stage process: for small compressions the chains contract by "buckling"
deformation whereas for larger compression the chains exhibit a collective
(almost uniform) bending deformation. Thus, the stiff polymer brush undergoes a
$2$-nd order phase transition of collective bond reorientation. The pressure,
required to keep the stiff brush at a given degree of compression, is thereby
significantly {\em smaller} than for an otherwise identical brush made of
entirely flexible polymer chains! While both the brush height and the chain
linear dimension in the $z$-direction perpendicular to the substrate increase
monotonically with increasing chain stiffness, lateral $(xy)$ chain linear
dimensions exhibit a maximum at intermediate chain stiffness. Increasing the
grafting density leads to a strong decrease of these lateral dimensions,
compatible with an exponential decay. Also the recovery kinetics after removal
of the compressing piston is studied, and found to follow  a power-law /
exponential decay with time.

A simple mean-field theoretical consideration, accounting for the
buckling/bending behavior of semi-flexible polymer brushes under compression, is
suggested.
\end{abstract}
\maketitle

\section{Introduction}
\label{sec_Intro}
Since several decades polymer brushes find abiding interest for various
applications, and have been studied very intensively by experiment, analytical
theory, and computer simulations \cite{1,2,3,4,5,6,7,8,9,10,11}. The structure
of these soft polymeric layers and their response to various external
perturbations depends in a delicate way on various control parameters such as
molecular weight, grafting density, quality of the solvent, and character of the
interactions between monomeric units and the substrate to which the
macromolecules are grafted. However, inspired by the early works \cite{12,13},
mostly brushes formed from completely flexible chains were considered, where the
elastic response is entirely entropic, due to the configurational degrees of
freedom of the tethered chain molecules. In such brushes, chains may be
nevertheless strongly stretched in the direction perpendicular to the substrate,
to avoid unfavorable monomer-monomer overlap in the case of sufficiently dense
grafting.

Effects of intrinsic chain stiffness, as are expected for semi-flexible polymers
\cite{14}, did find rather little attention \cite{15,16,17,18,19}, apart from
the case where one considers grafted semi-flexible chains in nematic solvents
\cite{20,21} and their liquid crystalline order \cite{22,23}. We also draw
attention to some related problems where rod-like molecules densely packed and
anchoring at substrates play a role, such as in Langmuir monolayers of
surfactants and related systems \cite{24,25,26,Schmid}. Experimentally, the
interactions between DNA-grafted colloids \cite{FKremer1,FKremer2}, or
poly(acrylic acid) brushes with variable rigidity \cite{FKremer3}, have been
comprehensively investigated by F.~Kremer and collaborators. In recent years,
semi-flexible polymers find increasing interest particularly in the context of
biophysical systems, such as cytoskeleton actin bundles \cite{27,28,29,30,31}
biofilaments \cite{32,33,34}, and also biobased polymer brushes were recently
created \cite{35}. Thus, a more comprehensive simulation study of brushes formed
from semi-flexible polymers would be desirable to theoretically clarify the
effect of chain stiffness on structure and dynamics of such systems, including
also their response to external perturbations, such as shear and compressive
forces.

As is well known, one important possible application of polymer brushes formed
from flexible polymers is their use as a lubricant \cite{4,6,10,36,37,38},
although the theoretical aspects of this response to shear and compression are
still under discussion \cite{39,40,41,42,43,44}. In our preliminary publication
\cite{45}, indeed, an anomalous response of brushes formed from rather stiff
chains was found, caused by an onset of orientational order parallel to the
substrate.

In the present paper we shall give a more complete account of this problem,
focusing mainly on the structure and static behavior of such stiff brushes
subject to compression. In the next Section \ref{sec_Model}, the model that is
studied will be introduced and the simulation method will be described. Section
\ref{sec_Structure} examines the parallel and perpendicular chain linear
dimensions in rigid brushes as a function of bending rigidity and grafting
density over a wide range, as well as the profiles of monomer density in the
direction perpendicular to the grafting substrate, and the end monomer
distribution. Then in Section \ref{sec_Compress} we elaborate the topic that was
already briefly addressed in our preliminary study, compression of such brushes
by a piston, presenting new results, including also the variation with grafting
density and chain length. The hysteresis associated with the transition towards
orientational order will also be investigated. Section \ref{sec_Theory} is
devoted to an attempt for theoretical consideration within the Mean-Field
Approximation of the brush response to deformation. A brief account on the
recovery dynamics of compressed stiff brushes is then given in Section
\ref{sec_Dynamics}. Our conclusions are summarized in section
\ref{sec_Summary}.

\section{Model and Simulation Methods}
\label{sec_Model}


We start from the standard model for polymer brushes formed from completely
flexible polymers \cite{3,46} and complement it by adding a bond angle
potential. This standard model is a bead-spring model in the continuum, where
bonded effective monomers interact with the well-known finitely extensible
non-linear elastic (FENE) potential \cite{47}
\begin{equation}\label{eq1}
V^{FENE} (r) = -0.5 kr_0^2 \ln [1-(r/r_o)^2], \; 0 < r < r_o ,
\end{equation}
where $r$ is the distance between the beads, and the spring constant $k$ as well
as the maximum distance $r_0$ between neighboring monomers will be specified
below. In addition, for any pairs of monomers (both bonded and non-bonded ones)
a truncated and shifted Lennard-Jones potential acts, the so-called
Weeks-Chandler-Andersen (WCA) potential \cite{48}
\begin{equation}\label{eq2}
V^{WCA}(r) = 4 \epsilon [(\sigma/r)^{12} - (\sigma /r)^6 + 1/4], \; r < r_c =
2^{1/6} \sigma ,
\end{equation}
while $V^{WCA} (r >r_c) = 0$. Note that this potential is cut off in its minimum
and both potential and forces are continuous for $r=r_c$.  We henceforth take
$\sigma = 1$ as our unit of length, and $\epsilon = T = 1$ (also Boltzmann's
constant $k_B$ is taken as unity, as usual). No explicit solvent molecules are
included, Eqs.~\ref{eq1}, \ref{eq2} describe interactions between effective
monomers and solvent molecules implicitly only, corresponding to the very good
solvent regime. The constants of the FENE potential are chosen as
\begin{equation}\label{eq3}
r_0=1.5 \sigma, \quad k = 30 \epsilon /\sigma ^2 ,
\end{equation}
so that the total potential between two subsequent monomers along the chain
($V^{FENE}(r) + V^{WCA}(r))$ has a minimum at about $r_{min}= 0.96 \sigma$.

The flexibility of the chains now is varied by introducing a bond angle
potential $V_b(\vartheta _{ijk})$ that depends on the angle $\vartheta _
{ijk}$
between the bonds between monomer pairs $(i,j)$ and $(j,k)$ respectively
\begin{equation}\label{eq4}
V_b(\vartheta _{ijk}) = \kappa _b [1-\cos (\vartheta_{ijk})]
\end{equation}

For rather stiff chains we have $\kappa_b \gg 1$ and hence Eq.~\ref{eq4} can be
approximated as
\begin{equation}\label{eq5}
V_b (\vartheta _{ijk}) \approx \frac {\kappa_b}{2} \vartheta^2_{ijk}\quad ,
\end{equation}
which shows that our polymer model can be viewed as a discretized version of the
well-known Kratky-Porod model \cite{49} of semiflexible chains (but, unlike the
Kratky-Porod model, the present model fully accounts for excluded volume effects
and is hence suitable for the study of rather dense systems as well). Note that
the effective persistence length of an isolated chain would be $\ell_p = \kappa
_b \sigma /k_BT = \kappa _b$; of course. Due to packing effects in dense
systems, local nematic order may arise as a result of interchain interactions
that lead to nematic short or long range order, and then the persistence length
can be appreciably larger (see e.g. \cite{50}). For small $\kappa _b$, the
persistence length is, in general, not a well defined quantity \cite{51}.

The chains are grafted at a planar $L \times L$ impenetrable surface, choosing a
square lattice arrangement of grafting sites, the lattice spacing chosen
according to the desired grafting density $\sigma _g$. Alternatively, also a
random arrangement of grafting sites has been considered, as discussed below.
The substrate surface exerts a repulsive interaction of a form analogous to
Eq.~\ref{eq3} on all monomers,
\begin{equation}\label{eq6}
V^{wall}(z) = 4 \epsilon _w [(\sigma _w/z)^{12} - (\sigma _w/z)^6 + \frac 1 4], \quad 0 <z<z_c = \sigma _w 2^{1/6}.
\end{equation}

Here $z$ is the distance of a monomer from the wall, and the parameters are
chosen as $\epsilon_w = \epsilon, \; \sigma _w = \sigma$. In $xy$-directions,
periodic boundary conditions are used, and in $z$-direction the simulation box
is closed with another repulsive wall (at distance $D$ from the grafting
surface) where a potential of the same type as Eq.~\ref{eq6} acts (simply $z$
needs to be replaced by $D-z$ on the right hand side of Eq.~\ref{eq6}). This
upper wall has no physical effect, as long as $D$ exceeds significantly the
height, $h_0$, of the unperturbed brush in equilibrium. However, we shall
consider also the effect of compressing the brush by reducing $D$ to values $D <
h_0$.

As an initial condition, all chains are put into a straight rod configuration,
choosing all angles $\theta _{ijk} =0$. Note that the first bond is constrained
to the $z$-direction, perpendicular to the grafting surface \footnote{This kind
of chemical grafting is referred to in the literature as {\em constraint
attachment} \cite{Arya} and shown to reveal quite unexpected effects of
stiffness in the context of poly-electrolyte colloids.}, so in the initial state
all chains are oriented along the $z$-axis. Then the chain configurations are
relaxed by standard Molecular Dynamics (MD) simulations, applying the Velocity
Verlet algorithm \cite{52} and a Langevin thermostat. Thus the coordinates
$\vec{r}_i(t)$ of the effective monomers evolve according to the Newton
equations of motion

\begin{equation}\label{eq7}
m \frac{d^2\vec{r}_i}{dt^2} = - \frac {\partial V^{tot}(\{\vec{r}_j\})}{\partial
\vec{r}_i} - \gamma \frac{d\vec{r}_i}{dt} + \vec{F}_i(t)\quad ,
\end{equation}
where the mass of effective monomers $m$ is chosen to be unity as well,
$V^{tot}$ is the total potential (containing the terms from the interactions
between monomers, Eqs.~\ref{eq1}, \ref{eq2}, \ref{eq4} and the wall-monomer
interaction, Eq.\ref{eq6}), $\gamma$ is a friction coefficient and
$\vec{R}_n(t)$ a random stochastic force. The latter satisfies the
fluctuation-dissipation relation \cite{52,53} ($\alpha, \beta$ denote the
Cartesian coordinates)
\begin{equation}\label{eq8}
\langle F_i^\alpha(t) F_j^\beta (t')\rangle \equiv 2 k_BT \gamma \delta _{ij} \delta _{\alpha \beta} \delta (t-t')
\end{equation}

The friction coefficient $\gamma$ is chosen as $\gamma = 0.25$. This choice
ensures efficient equilibration. We also note that with our choice of units, the
MD time unit is unity as well, $\tau _{MD} = \sigma (m/\epsilon)^{1/2} =1$.
In our simulations, $\kappa _b$ is varied from $\kappa_b = 0$ (flexible chains)
up to $\kappa _b = 50$, and chain lengths from $N=10$ to $N=80$, so cases are
included where the contour length and the persistence length are of the same
order of magnitude. Grafting densities are varied from $\sigma _g=0.0625$ (the
``mushroom regime'' \cite{11,12,13}) to $\sigma _g=1$ (in the latter case
monomer density inside of the brush corresponds to a concentrated solution). The
number of chains was typically chosen to be $\mathcal{N}_{chain}=128$, but in a
few cases the lateral box linear dimensions were doubled (so that
$\mathcal{N}_{chain}=512$) so as to verify that finite size effects are still
negligibly small.

\begin{figure}[htb]
\vspace{0.5cm}
\includegraphics[scale=0.29]{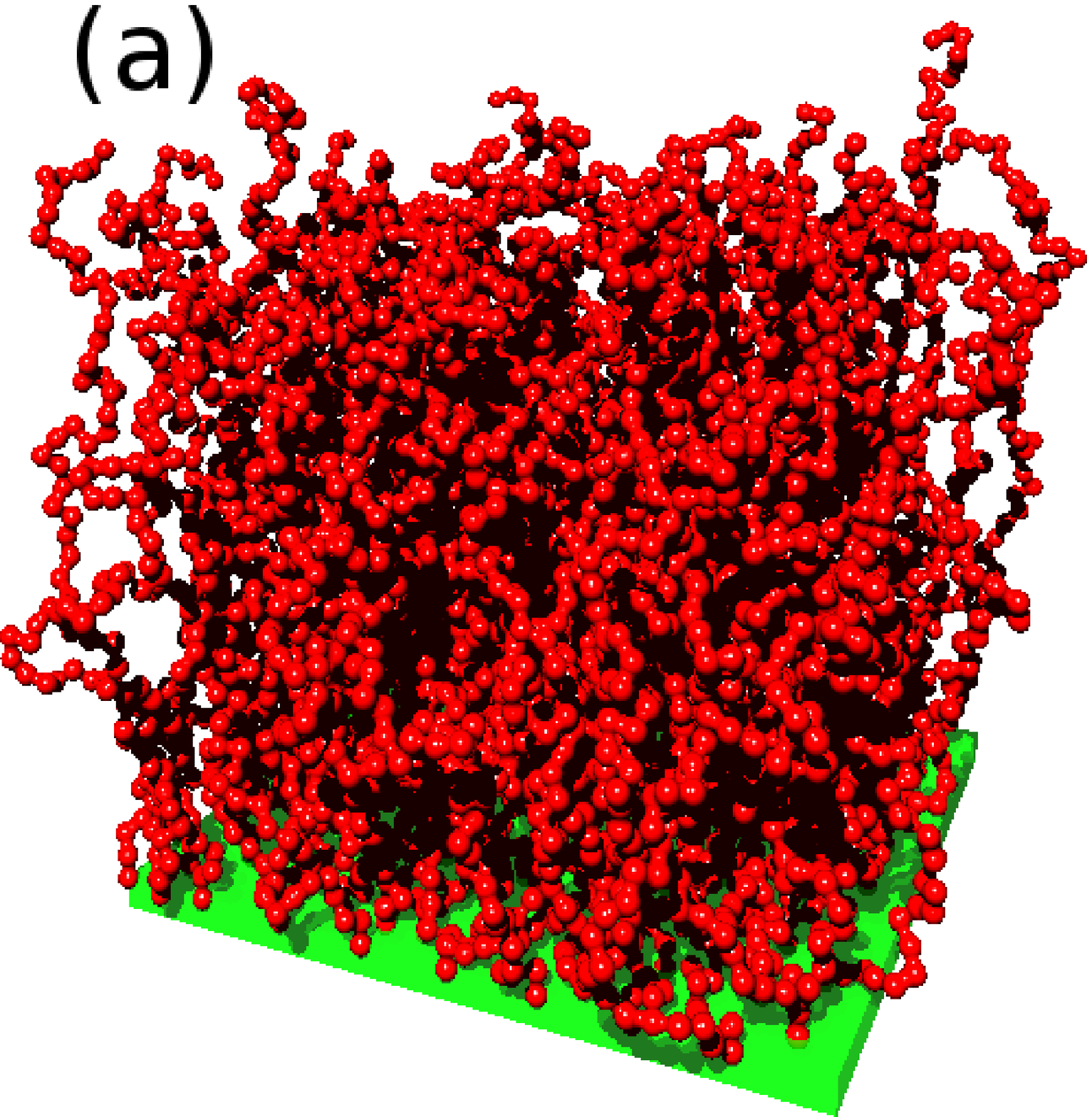}
\hspace{0.7cm}
\includegraphics[scale=0.29]{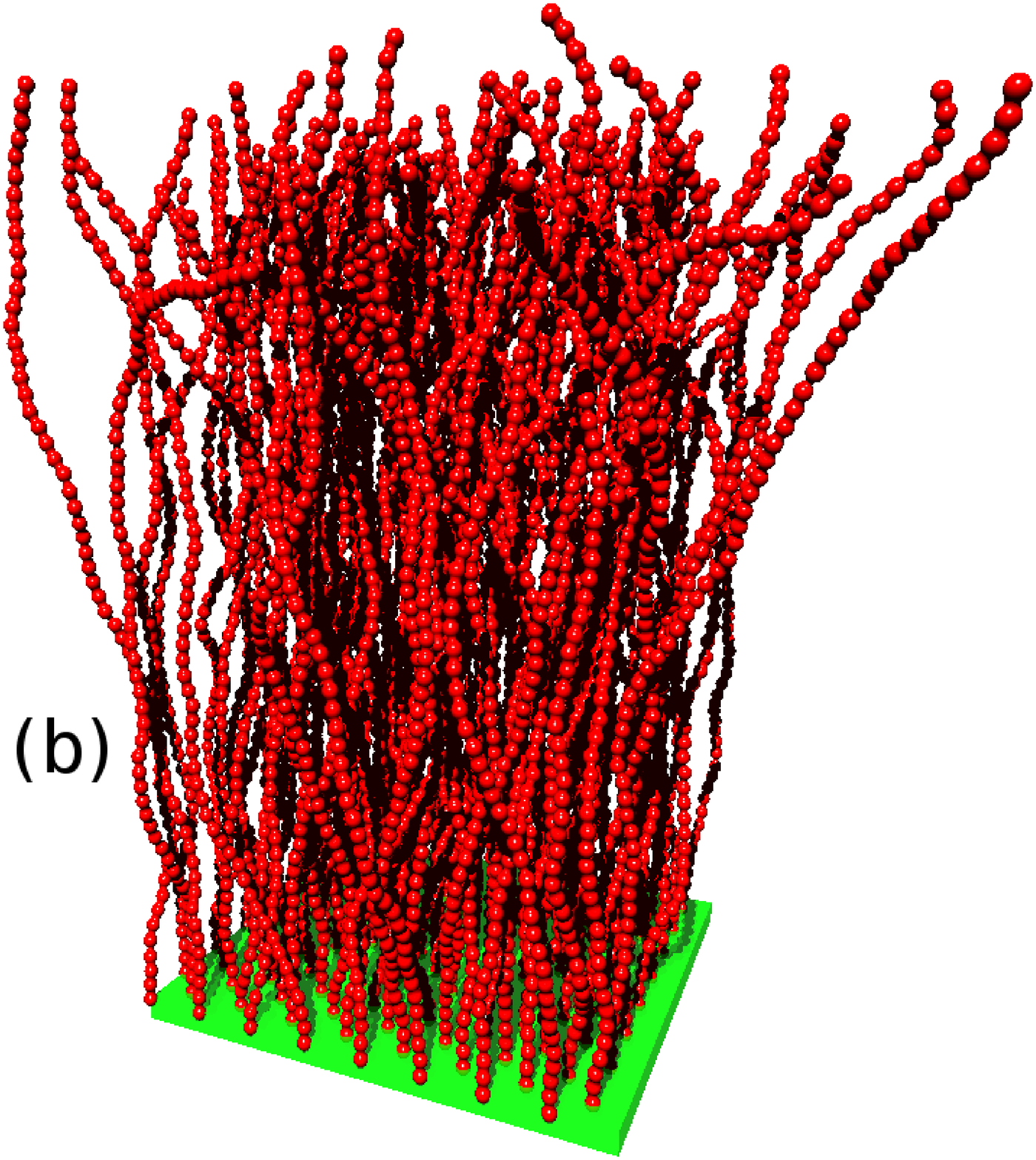}
\caption{\label{fig1} Snapshot pictures of polymer brushes for the
case $L = 32$, $\sigma _g = 0.125$, $N=60$, and two choices of $\kappa_b$: \;
$\kappa_b = 0$ (fully flexible chains) (a), $\kappa_b= 50$ (semi-rigid
chains) (b). The distance between observer and objects is the same in both
cases.}
\end{figure}
As an example, Fig.~\ref{fig1} shows typical snapshot pictures for a
well-equilibrated brush at a grafting density $\sigma _g = 0.125$ and $N=60$ for
two choices of $\kappa_b$, $\kappa_b=0,\;50$. One can clearly see that for the
flexible case the chain conformation is locally rather random, the overall
stretching of the chains in $z$-direction becomes visible on larger scales only.
In contrast, the stiff chains show at most long wavelength transverse
undulations, like a bunch of pliant rods. This behavior is exactly what one
expects to occur for worm-like chains, which hence are well modeled by our
simulation. In the next section we shall analyze the properties of these brushes
in more detail.

\section{Structure of semiflexible polymer brushes}
\label{sec_Structure}

\begin{figure}[htb]
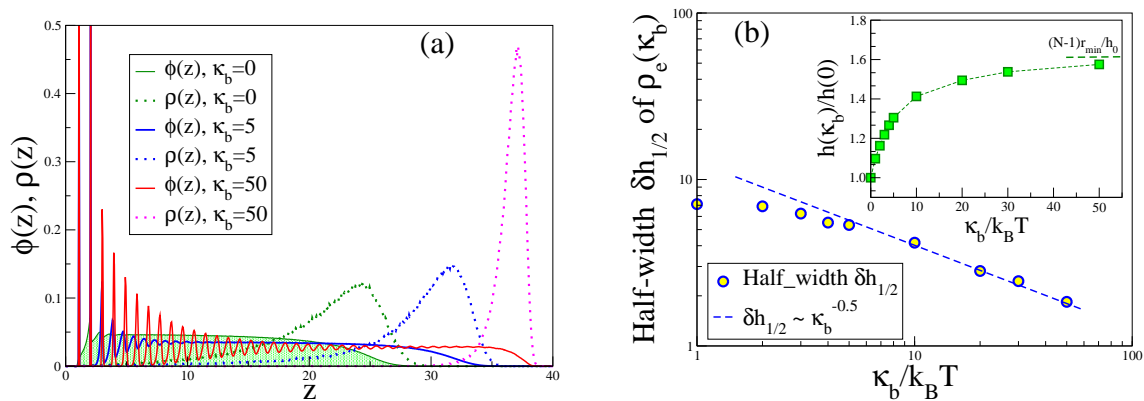

\vspace{0.5cm}
\includegraphics[scale=0.30]{phi_z_N40_s_g_0.25.eps}
\hspace{0.7cm}
\includegraphics[scale=0.30]{h_k.eps}

\caption{\label{fig2} (a) Density distribution $\phi(z)$ of the effective
monomers in the brush, and distribution of chain ends, $\rho(z)$, plotted vs.
the distance $z$ from the grafting surface, for $N = 40,\; \sigma_g = 0.25$, and
three choices of $\kappa_b$, as indicated. The density profile for $\kappa_b =
0$ is indicated by shaded area. Note the normalizations $\int \limits _0^\infty
\phi (z) dz=1$, $\int \limits _0^\infty \rho (z) dz=1$. (b) Half-width of
$\rho(\kappa_b)$ vs $\kappa$ for $N = 40$ and $\sigma_g = 0.25$, shown as a
$\log$-$\log$ plot. The inset shows the relative increase of the brush height
with $\kappa_b$. The asymptotic value $h(\kappa_b \to \infty)/h_0 =
(N-1)r_{min}/h_0$ is indicated by the dashed horizontal line.}
\end{figure}

We start with a discussion of the monomer density profile, $\phi(z)$, and the 
distribution of free ends, $\rho(z)$, for a typical choice of $N$ and
$\sigma_g$, $N=40$ and $\sigma_g =0.25$, varying $\kappa_b$ (Fig.~\ref{fig2}a).

One notes that near the wall there is always a pronounced layering effect, which
strongly increases with $\kappa_b$, although the density of the stiffer chains
is smaller in the center of the brush (near $z=h/2$, as a rough measure of brush
height $h$, we may take the inflection point of the profile $\phi(z)$ near its
final decay for large $z$). This density decrease with increasing $\kappa_b$
must occur since the stiffer chains are more stretched and the brush height
$h_0$ increases with increasing $\kappa_b$ up to its maximum value while the
number of monomers stays constant. This effect that stiffer chains exhibit more
pronounced layering than flexible chains has already been pointed out for a
different model \cite{19}. Note that the actual value of the density is
$N\phi(z)$ and hence about $0.44$ for $\kappa_b =0,\;0.33$ for $\kappa_b = 5$,
and $0.26$ for $\kappa_b = 50$. These densities are much smaller than melt
densities, indicating a rather unexpected high degree of order for the
semi-flexible brushes near the wall.  Fig.~\ref{fig2}b shows that the
distribution of chain ends becomes sharper with increasing chain stiffness. The
half-width $\delta h_{1/2}$ scales as $\delta h_{1/2} \propto \kappa_b^{-0.5}$
for large $\kappa_b$ (i.e., $(\delta h_{1/2})^2$ scales inversely with the
persistent length).

\begin{figure}[htb]
\vspace{0.7cm}
\includegraphics[scale=0.30]{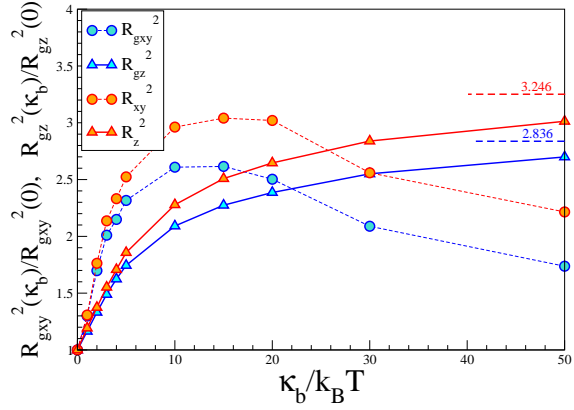}

\caption{\label{fig3} Normalized mean square gyration radii components and
end-to-end distance components $R^2_{gz}(\kappa_b)/R^2_{gz}(0), \;
R_z^2(\kappa_b)/R_z^2(0)$, $R^2_{gxy}(\kappa)/R^2_{gxy}(0), \;
R^2_{xy}(\kappa_b)/R^2_{xy}(0)$ in the directions perpendicular $(z)$ and
parallel $(xy)$ to the grafting surface, plotted vs. $\kappa_b$ for $N = 40$
and $\sigma_g = 0.25$. Dashed horizontal lines indicate the asymptotic ratio
values for $\kappa_b \to \infty$ whereby $R_z(\kappa_b \to \infty) =
(N-1)r_{min}$ and $R_{gz}(\kappa_b \to \infty) = \frac{1}{12}\left[
r_{min}(N-1)\right]^2$. }
\end{figure}

The height of the brush $h(\kappa_b)$ saturates for $\kappa_b \rightarrow
\infty$ at its maximum value of fully stretched chains, $h_{max}=0.96\; (N-1)$,
see the inset of Fig.~\ref{fig2}b, so the distribution of chain ends becomes a
delta-function! Fig.~\ref{fig3} presents the chain linear dimensions as
functions of $\kappa_b$, for $N =  40$ and $\sigma _g =0.25$. As expected, the
chains get more stretched as the polymers get stiffer. Interestingly, also the
linear dimensions in $xy$-directions parallel to the wall initially increase
even stronger than the linear dimensions in $z$-direction. Due to the increased
stretching, the density in the brush has decreased, and hence there is more
space for the monomers to occupy. In particular, the large magnitude of this
effect is surprising. It is also evident from the snapshot - Fig.~\ref{fig1}b.
We also note that the linear dimensions $R_{gxy}^2, R_{xy}^2$ reach their
maximum in the range $10 \leq \kappa_b \leq 20$, i.e., a range where the contour
length is just a few times larger than the persistence length. Note that for a
single semi-flexible chain one predicts that transverse fluctuations of monomers
in the rod-like limit scale as \cite{54}
\begin{equation}\label{eq9}
\langle (\delta \vec{r}_{xy})^2\rangle \propto L^3/\ell_ p
\end{equation}

Thus, Fig.~\ref{fig3} suggests that a similar decrease of the linear dimensions
in the $xy$-direction (transverse to the average orientation of the rod-like
grafted chains in the brush, which is the $z$-direction) occurs also in fairly
dense brushes. However, much more data (for still larger values of $\kappa_b$)
would be required to test whether Eq.~\ref{eq9} holds here quantitatively. Note
that for $\kappa_b = 50$ the $xy$-components of the mean square gyration radius
are still larger than their counterparts for flexible chains, although the brush
height is close to its maximum value. For $\kappa_b \rightarrow \infty$, these
$xy$-components must tend to zero.

\begin{figure}[htb]
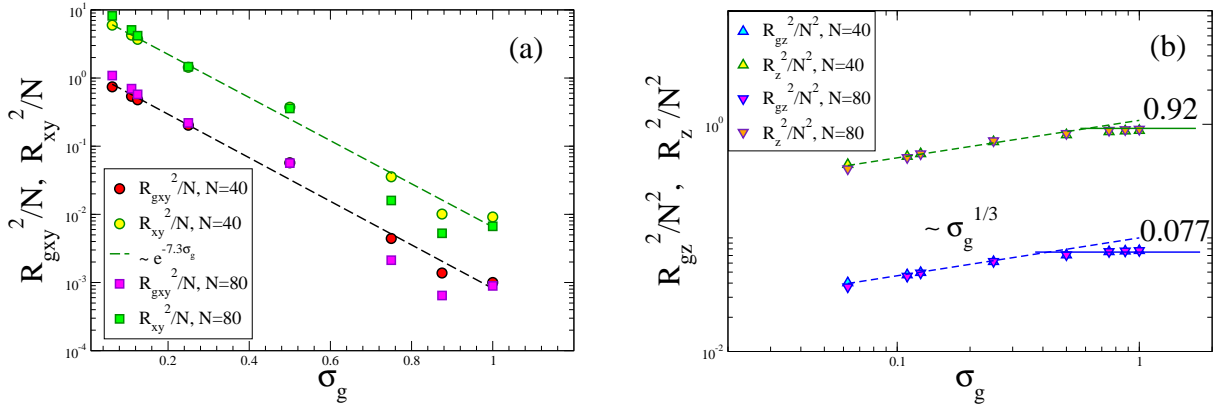

\vspace{0.7cm}
\includegraphics[scale=0.30]{Rg_sigma_par.eps}
\hspace{0.7cm}
\includegraphics[scale=0.30]{Rg_sigma.eps}

\caption{\label{fig4} (a) Plot of $R^2_{gxy}/N$
and $R^2_{xy}/N$ vs. $\sigma_g$ for $\kappa_b=20$ and two choices of N, N = 40
and N = 80. Note the logarithmic ordinate scale; the straight lines indicate an
empirical exponential relation, $\exp (-7.3 \sigma_g)$. (b) Log-log plot of
$R_{gz}^2/N^2$ and $R_z^2/N^2$ vs. $\sigma_g$. Straight lines indicate
variations proportional to $\sigma_g ^{1/3}$. All data refer to $\kappa_b = 20$.
Both data for $N = 40$ and $N = 80$ are shown. Horizontal straight lines in (b)
indicate the asymptotic results when the chains are fully stretched out like a
straight rod along the $z$-axis, i.e., $R_z^2/N^2=r_{min}^2 \approx 0.92$ and
$R^2_{gz}/N^2 = r_{min}^2/12=0.077$, respectively.}
\end{figure}

Fig.~\ref{fig4} studies the variations of chain linear dimensions with
$\sigma_g$. While for flexible chains one knows that in the semi-dilute limit
$R_z \propto \sigma_g^{1/3}N$, and hence $R_{gz}^2 \propto \sigma_g^{2/3}N^2$,
that is, our data are compatible with a variation $R_{gz}^2 \propto  N^2$, but
the variation with $\sigma _g$ is distinctly weaker. The values reached for
these large values of $\sigma_g$ indicate that the chains are stretched out like
straight rods, since in this rod limit we simply have $R_z^2/N^2 \approx 0.92$
and $R_{gz}^2/N^2 = r_{min}^2/12 = 0.077$, respectively. Note that for $\sigma_g
\geq 0.5$ thus a saturation is reached.

Similarly, for flexible chains one would have $R_{gxy}^2 \propto
\sigma_g^{-1/6}N$; however, the present data suggest a much more rapid decay of
$R^2_{gxy}$ with $\sigma_g$ than this weak power law. Empirically we find that
the decrease is compatible with an exponential decay: Unfortunately, we do not
have any explanation for this puzzling behavior. However, it should be remarked
that for semi-flexible brushes the regime where the crossover from semi-dilute
brushes to mushroom-like behavior occurs is found for much smaller grafting
densities than for brushes formed from flexible chains of the same length. This
fact simply results from the observation that the size of a chain in dilute
solution scales as $R \propto \ell_b N^{3/5}$ for flexible chains, while for
semi-flexible chains the size is much larger \cite{55}, $R \propto \ell_p^{1/5}
\ell_b^{4/5} N^{3/5}$ for very long chains $(N > (\ell_p/\ell _b)^3$) and $R
\propto (\ell_p \ell_b)^{1/2} N^{1/2}$ for chains with $\ell_p / \ell_b < N <
(\ell_p/\ell_b)^3$; the bond length is denoted as $\ell_b$ in this scaling
description, and can be identified with the distance $r_{min}$ of the minimum of
our effective potential between neighboring monomers in our model. Since we use
here rather stiff chains of medium length, we are in the regime $N
<(\ell_p/\ell_b)^3$ with our simulations always. Hence the crossover from
mushrooms to semi-dilute brushes occurs for $\sigma _g = \sigma _g^*$ given by
\begin{equation}\label{eq10}
\sigma _g^* \approx R^{-2} = \ell_p^{-1} \ell_b^{-1} N^{-1}\quad .
\end{equation}

The scaling behavior typical for semi-flexible semi-dilute brushes hence is
expected for $\sigma_g ^* \ll \sigma _g \ll 100 \sigma _g^*$. This regime
clearly is not explored in our simulations, which rather address the regime from
concentrated solutions to melt densities.

\section{Response of Semiflexible Brushes to Compression - Simulation results}
\label{sec_Compress}

\begin{figure}
\vspace{0.7cm}
\includegraphics[scale=0.20]{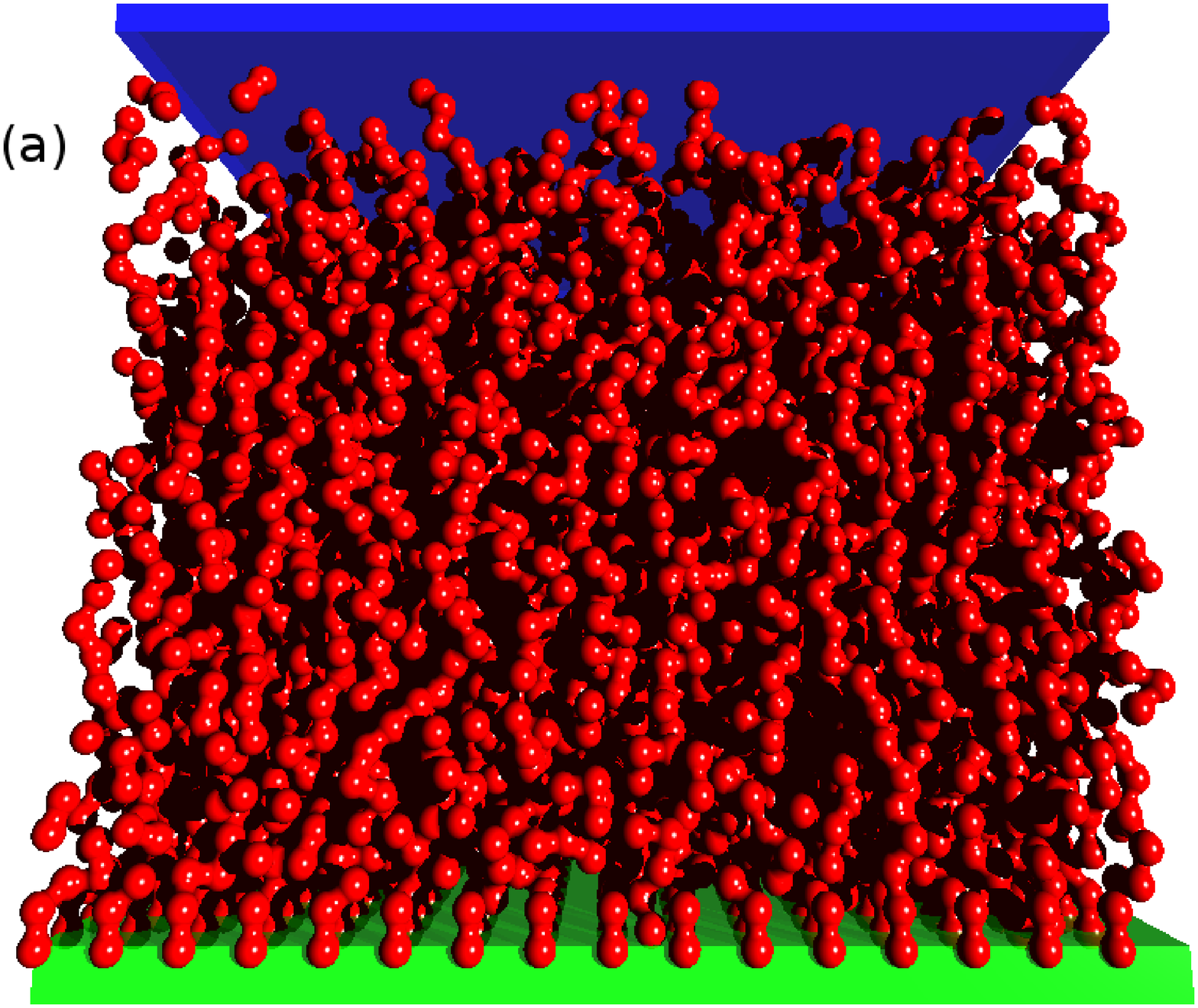}
\hspace{0.7cm}
\includegraphics[scale=0.20]{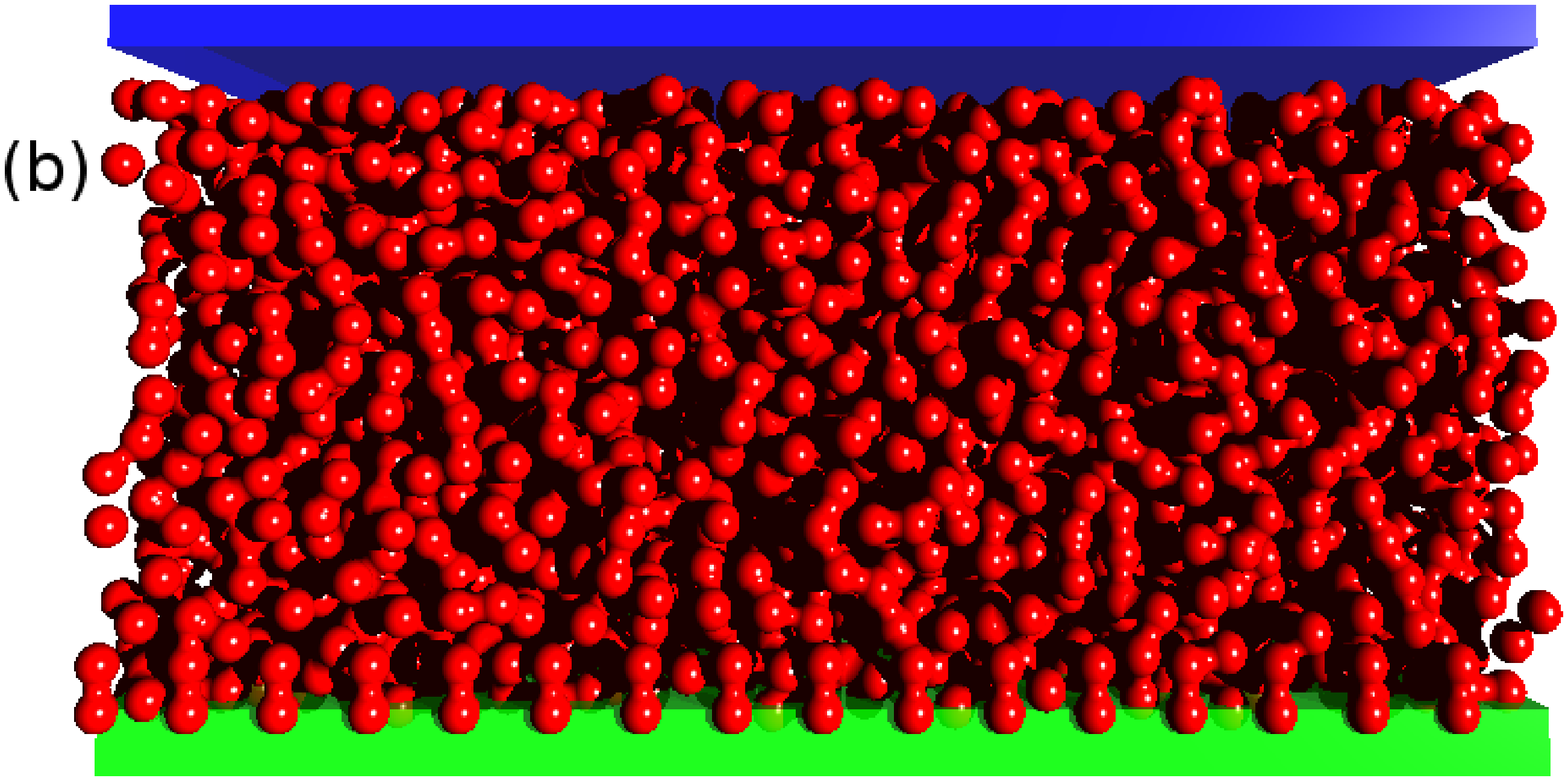}
\vspace{0.7cm}

\includegraphics[scale=0.20]{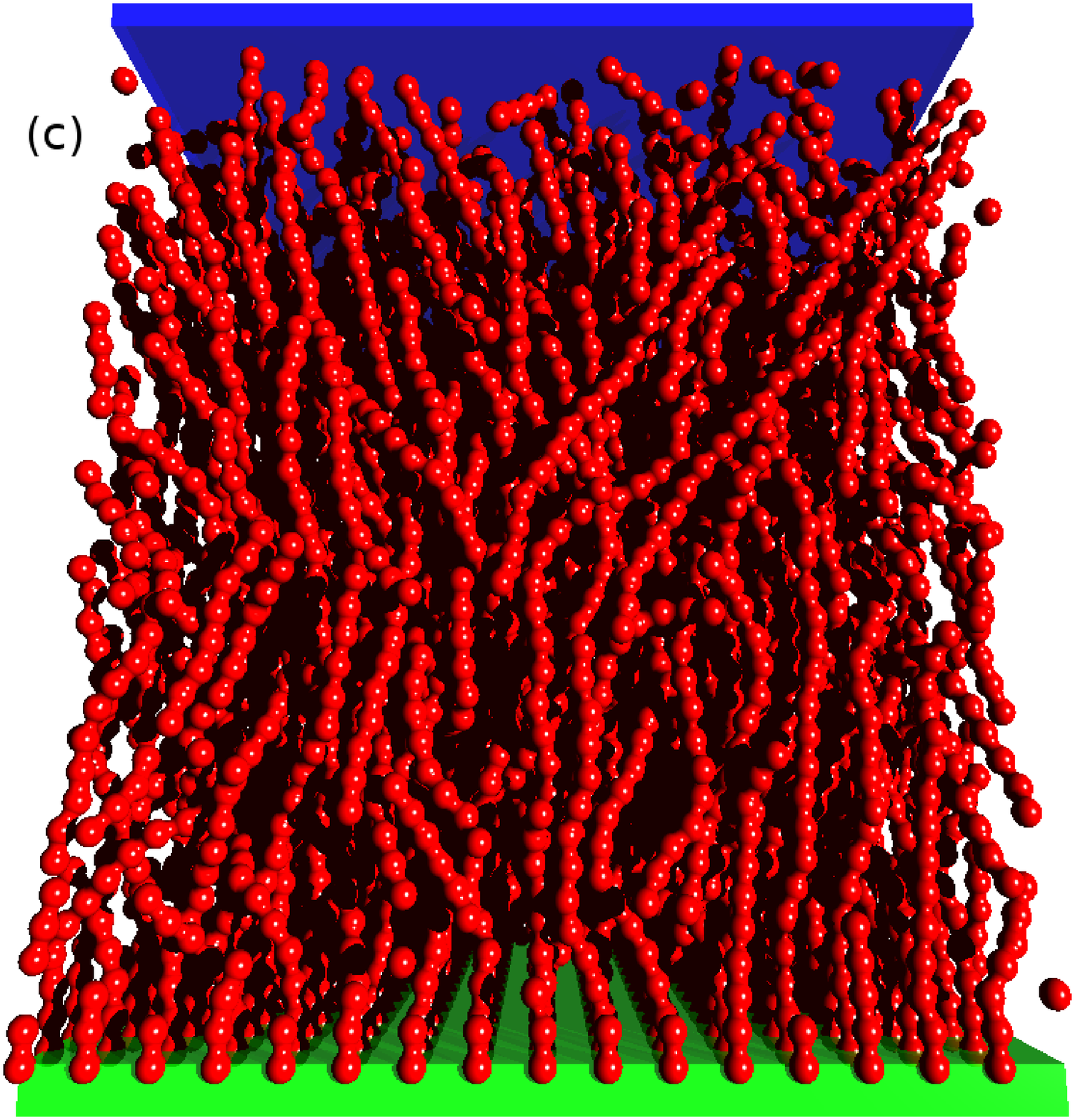}
\hspace{0.7cm}
\includegraphics[scale=0.20]{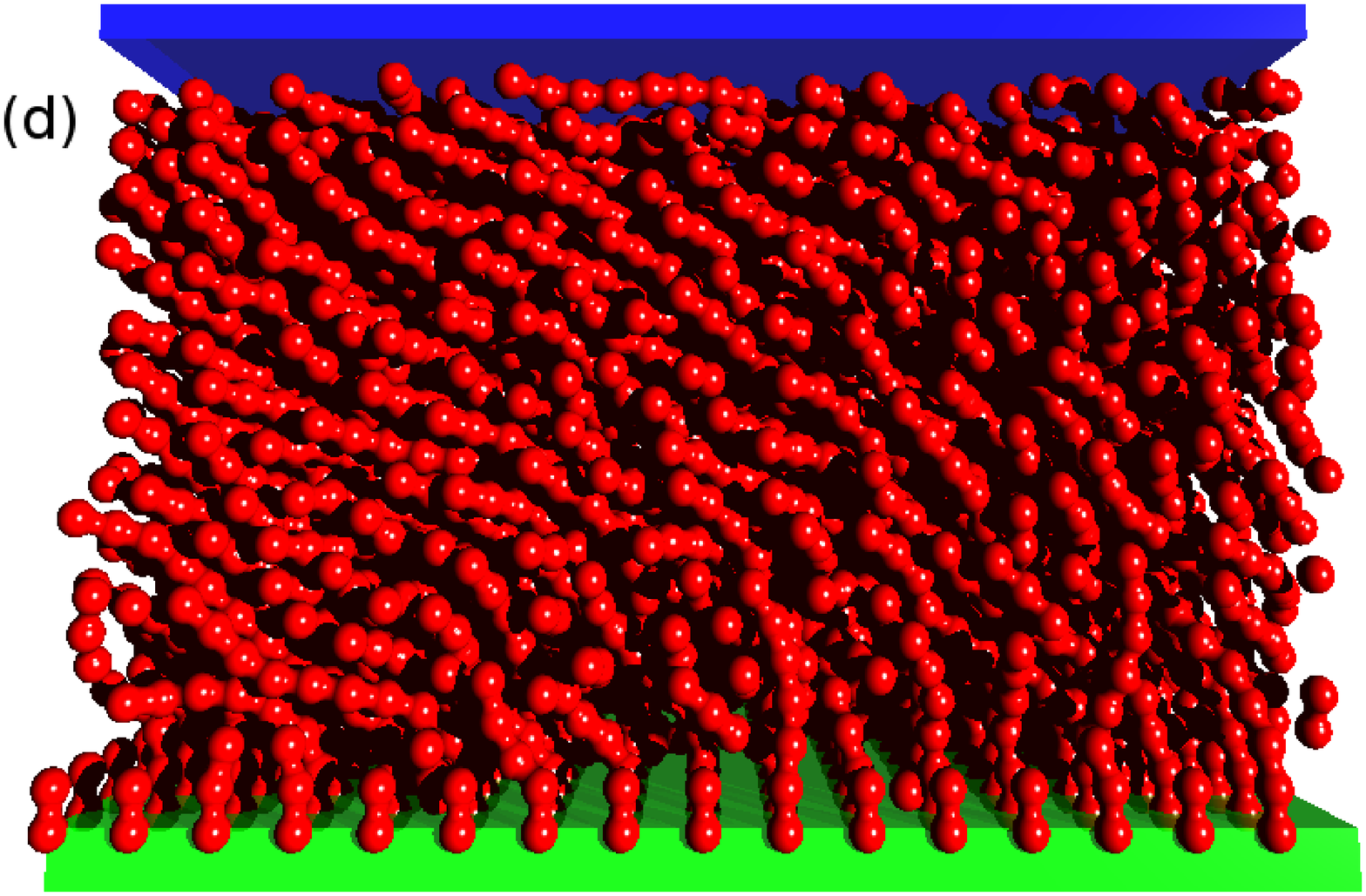}

\caption{\label{fig5} Snapshot pictures of polymer brushes for the case of L = 
32, N = 40, $\sigma_g = 0.25$, and two choices of $\kappa_b$, $\kappa_b =0$ 
(a,b) and $\kappa_b =20$ (c,d), applying a compression characterized by $D/h_0 = 
0.9$ (a,c) and $D/h_0=0.5$ (b,d). } 
\end{figure}

\begin{figure}[htb]
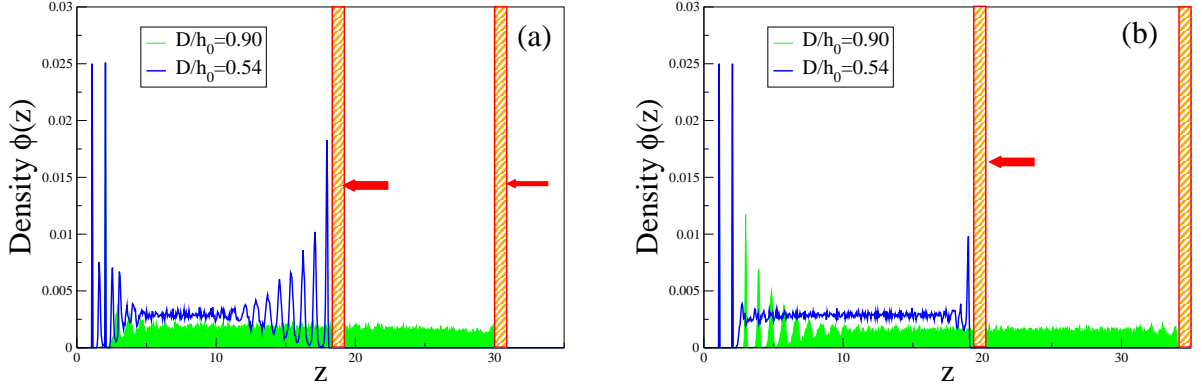

\vspace{0.7cm}
\includegraphics[scale=0.30]{histo_N40_s0.5_flex.eps}
\hspace{0.7cm}
\includegraphics[scale=0.30]{histo_N40_s0.5_stiff.eps}

\caption{\label{fig6} Density distribution of the effective monomers in the 
brush, $\phi (z)$, plotted vs distance $z$ from the grafting surface, for 
$N=40$, $\sigma =0.5$, for two choices of $D/h_0$, namely, $D/h_0=0.9$  and 
$D/h_0=0.54$. Two choices of $\kappa_b$ are included namely fully flexible 
chains $(\kappa_b=0$, case (a), and semiflexible ones $(\kappa_b = 20)$, case 
(b).}
\end{figure}

In this section we consider the change of the state of the brush when it is
compressed by a flat structureless piston parallel to the grafting surface that
is brought to a height $D$ (= distance from the grafting surface) less than the
height $h_0$ of the free brush. Fig.~\ref{fig5} shows a few representative 
snapshot pictures and Fig.~\ref{fig6} corresponding density profiles. One can 
see that for the
flexible chain the density increase caused by the compression leads to a more
pronounced layering of the monomers near the grafting surface; also at the
compressing upper surface the parabolic decay of the density profile found in
the free brush is replaced by the density oscillations in the compressed brush.
For very strong compression the layering at the compressing wall is even more
pronounced than near the grafting surface, where the brush conformation is more
constrained due to the condition that the first bond of each grafted chain must
be oriented in the $z$-direction perpendicular to the grafting surface. For the 
semi-flexible grafted chains the compression has the effect to clearly reduce 
the extent in $z$-direction over which layering can be seen, in comparison with 
the corresponding uncompressed brush.

\begin{figure}[htb]
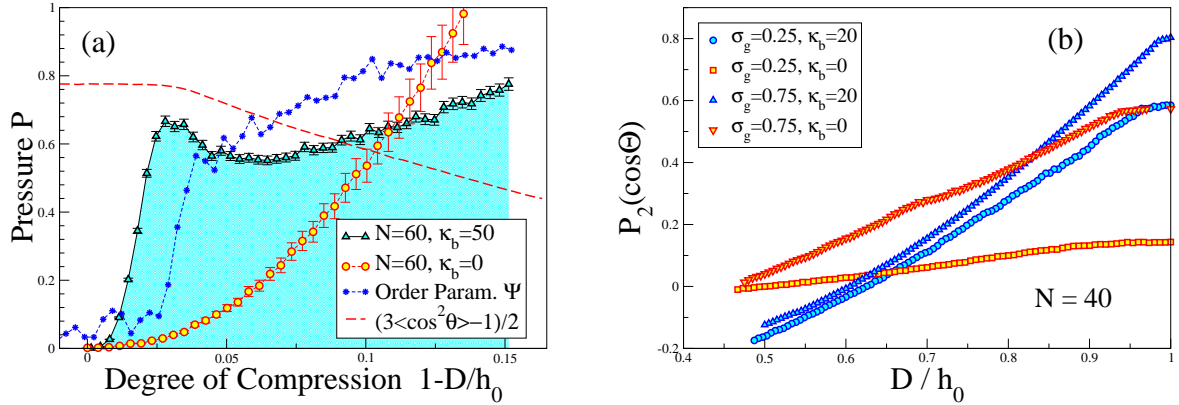

\vspace{0.7cm}
\includegraphics[scale = 0.30]{P_z_N60_s_0.5.eps}
\hspace{0.7cm}
\includegraphics[scale=0.30]{Legendre.eps}

\caption{\label{fig7} (a) Normal pressure $\sigma P/k_BT$ for compressed brushes
with chain length $N = 60$ and grafting density $\sigma _g =0.5$ plotted vs. the
degree of compression, $1-D/h_0$. The brush contains $N_{ch}=512$ chains. Both a
flexible brush $(\kappa_b=0$) and a brush formed from rather stiff chains
($\kappa_b=50$) are included; for the latter also the order parameter $\Psi$
\{Eq.\ref{eq10}\} and the orientational order parameter $P_2(\cos \theta)$
\{Eq.~\ref{eq11}\} are shown, as indicated. (b) $P_2(\cos \theta)$ plotted vs.
$D/h_0$ for the case $N=40$. Two values of the grafting density are shown
$(\sigma_g = 0.25$ and $0.75$, respectively) for both flexible $(\kappa_b=0$)
and semi-flexible $(\kappa_b=20$) chains.}
\end{figure}

The snapshot pictures (Fig.~\ref{fig5}) give a qualitative interpretation for
this surprising behavior: While for the flexible brush the chains get uniformly
compressed and the density increases gradually, the rather stiff chains exhibit
a collective bending, and the interplay of the bending of the stiff chains and
the dense packing of the effective monomers destroys the periodicity of the
local density that we see in the uncompressed case. It is found that the onset
of collective chain bending does not start immediately when the distance $D$
between the compressing piston surface and the grafting substrate become equal
to the uncompressed brush height $h_0$, but only when the degree of compression
$1-D/h_0$ exceeds a threshold value. This fact is clearly recognized
(Fig.~\ref{fig7}) when we study the normalized pressure $\sigma P/k_BT$ (note
that the normal pressure $P$, i.e. the diagonal component of the pressure tensor
$p_{zz}$, is easily sampled using the virial theorem \cite{52}). For a free,
uncompressed brush in thermal equilibrium $P=0$, of course. For very small
compression, $1-D/h_0 \leq 0.005$, the pressure is almost immeasurably small.
Note, however, that there is some arbitrariness in defining $h_0$ exactly: here
we have defined $h_0$ from the condition that $\phi(z)$ \{Fig.~\ref{fig2}a\} has
decreased to about $1\%$ of its value in the flat region of $\phi(z)$ where the
oscillations of $\phi(z)$ have just decayed. Note that the definition in terms
of the first moment $\langle z \rangle$ of the density profile, $h_0 = 8 \langle
z \rangle /3$, that is often used for flexible brushes and holds for the
parabolic profile of the self-consistent field theory \cite{1,10,11}, is not
useful for rather dense semi-flexible brushes, and hence not used here. 

\begin{figure}[htb]
 \vspace{0.7cm}
 \includegraphics[scale = 0.30, angle=270]{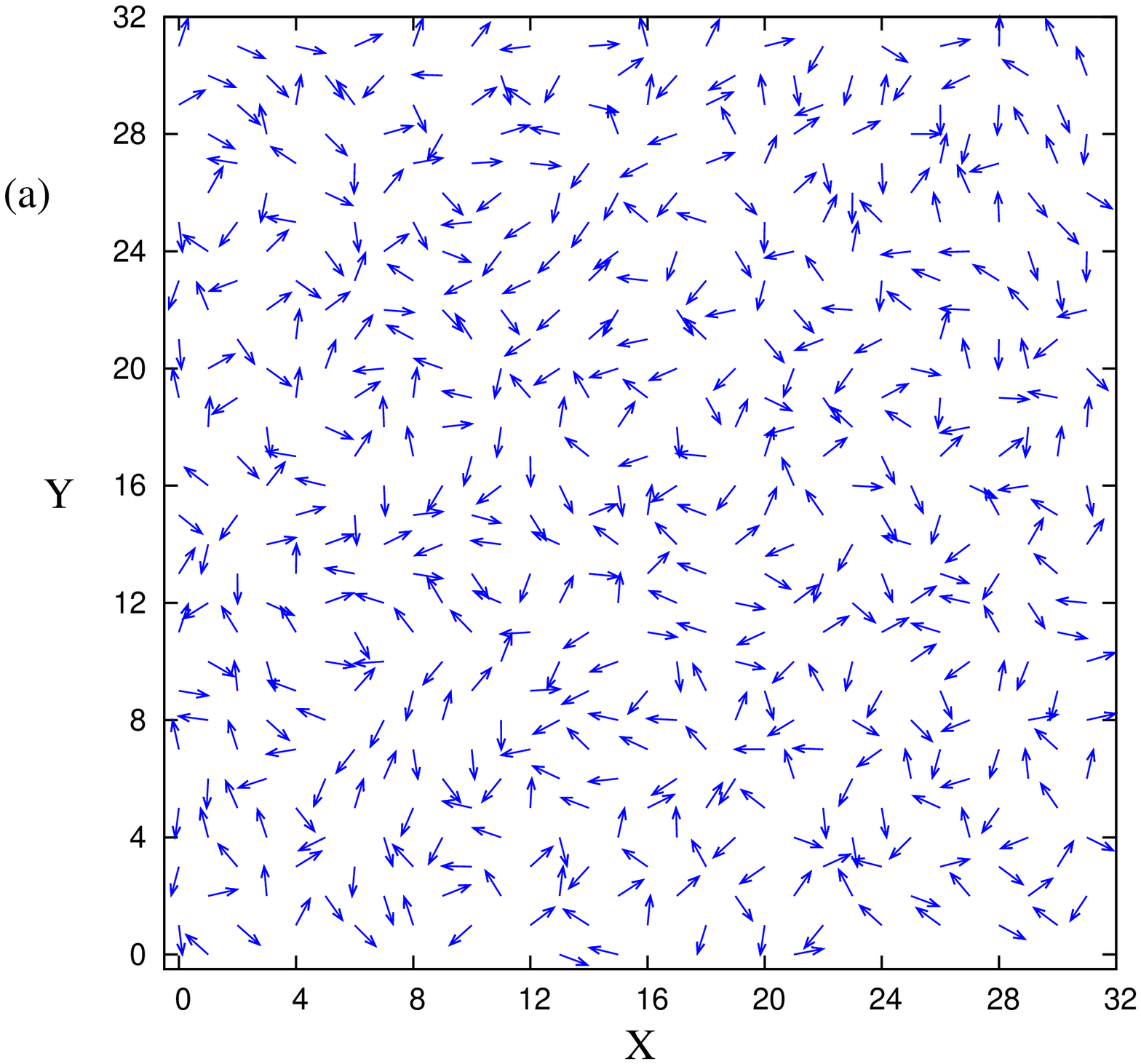}
 \hspace{-0.7cm}
 \includegraphics[scale = 0.30, angle = 270]{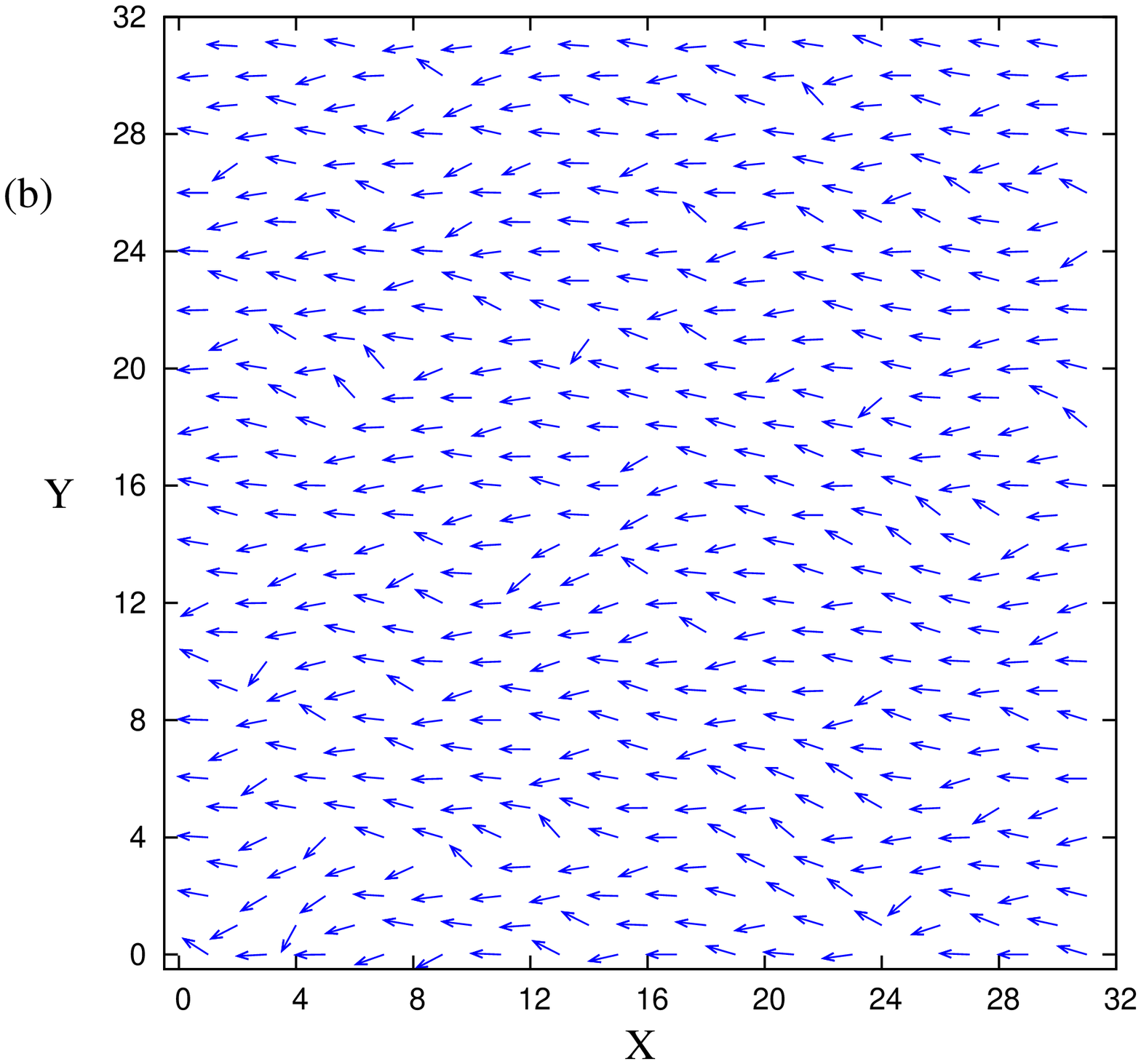}
 
 \caption{Orientation of the last bonds in a polymer brush, represented by unit 
vectors in the $x,y$-plane for different degrees of compression $1-D/h_0$: (a)
no compression, orientational disorder; (b) $25\%$ compression, orientational 
order. Here $N=40,\; \sigma_g = 0.5,\; \kappa_b = 50$, and the system size is
 $L\times L = 32^2$ with $N_{ch}=512$. \label{fig_vec} }
\end{figure}

While for flexible brushes, $(\kappa_b =0)$, the increase of the pressure is 
completely gradual for the full range of compressions, and can be described by a 
power law (roughly $P \propto D^{-3}$ for $\sigma \ll D \ll h_0$, as will be
discussed below), the variation of the pressure for stiff chains is very
different: there occurs a much faster rise up to a maximum, then the pressure
decreases again to a shallow minimum, before a slow increase of the pressure
occurs again. At large compressions the pressure in semi-flexible brushes
clearly is distinctly {\em smaller} than in their flexible counterparts! The
pressure maximum is due to the onset of collective orientational ordering of the
chains in a particular direction in the $xy$-plane. A convenient way to measure
this ordering in terms of the unit vectors $\vec{u}_k$ of the projections of the
last bond vector of each chain into the $xy$-plane is
\begin{equation}\label{eq10a}
\Psi = \left \langle \frac{1}{N_{ch}}\sqrt{ \sum \limits _{k=1} ^{N_{ch}}
u_{k,x}^2 + \sum \limits _{k=1}^{ N_{ch}} u_{k,y}^2}  \right \rangle
\end{equation}
as illustrated by the snapshots in Fig.~\ref{fig_vec}a,b.

If these unit vectors $\vec{u}_k$ are randomly oriented in the $xy$-plane, we 
obviously have $\Psi = 1/\sqrt{N_{ch}}$, and this is what is observed for zero 
(or very small) compression. This nonzero-plateau of $\Psi$ at zero compression 
is hence just a trivial finite size effect. On the other hand, if spontaneous
symmetry-breaking occurs, and all unit vectors are parallel to each other, we
would have $\Psi = 1$. Such definitions of root mean square order parameters are
well known from simulation studies of phase transitions in isotropic
ferromagnets \cite{56}. This gradual onset of ordering also shows up in the
average angle that bonds make with the $z$-direction. Thus Fig.~\ref{fig7}b
includes also data on $P_2(\cos\theta)$ defined as usual by
\begin{equation}\label{eq11}
P_2(\cos \theta ) = \frac 1 2 (3 \langle \cos ^2 \theta\rangle -1)
\end{equation}
While in the initial period of compression, up to the pressure maximum,
$P_2(\cos\theta)$ stays essentially constant, then a gradual decline of
$P_2(\cos \theta)$ sets in. Recall that $P_2(\cos\theta) = 1$ for rods oriented
along the $z$-axis while $P_2(\cos\theta) = -\frac{1}{2}$ for rods oriented
parallel to the $xy$-plane. However, this smooth behavior of the order
parameters $\Psi $ and $P_2 (\cos \theta)$, indicative of a continuous
(second-order) phase transition, is at variance with the observation of a loop
(hysteresis) of the pressure in its variation with the compression. Such loops
normally would be associated with a
\begin{figure}[htb]
\includegraphics[scale=0.30]{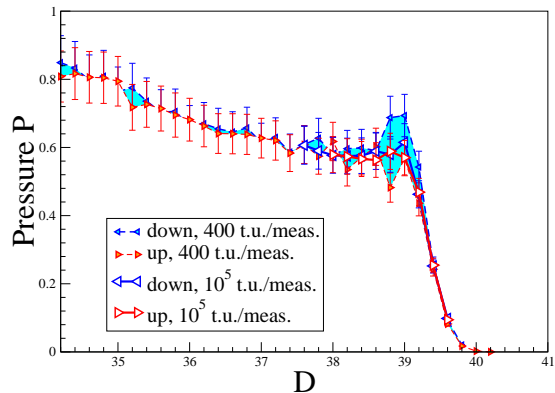}

\caption{\label{fig8} Normal pressure $\sigma P/k_BT$ plotted vs. $D$ for the 
case $N=40$, $\kappa_b = 50$ and $\sigma _g = 0.50$, comparing runs where the 
pressure is stepwise increased and decreased again. In a short run, at each 
value of $D$ the system is held $400$ MD time units whereas in a long run, the 
equilibration time at every state was $25$ times longer. The variance of
measured values is also plotted as error bars.}
\end{figure}
discontinuous (first order) phase transition. But a more careful consideration 
of equilibration reveals, (Fig.~\ref{fig8}), that although some hysteresis 
between runs, where the compression is stepwise increased or decreased, is 
actually observed, it must be interpreted as an observation time effect: the 
amount of the hysteresis is almost completely gone, if the observation times are 
chosen $25$ times larger! Of course, we expect that close to a second order
transition relaxation times in the system become very long, due to ``critical 
slowing down'' \cite{56,57}. Thus, hysteresis is also known to occur for 
simulations of second order transitions \cite{56}, if the observation times at 
the individual state points near the transition are chosen too short. While for 
most systems this problem is not important in practice \cite{56}, here the 
situation differs because densely arranged stiff polymers are slow objects: the 
$xy$-like order of the last bond must be shared by a related order of the inner 
bonds of each chain as well. Figs.~\ref{fig7}, \ref{fig8} focus on the behavior 
in the vicinity of the phase transition only.

It turns out, (Fig.~\ref{fig7}b), that with strong compression the orientation
of the bonds, which in Figs.~\ref{fig7},~\ref{fig8} is still mostly along the
$z$-direction, can be changed to a preference of $x,y$ direction: then, $P_2
(\cos \theta) < 0$ for semi-flexible brushes. In contrast, flexible brushes at
similar compressions, $(D/h_0 \approx 0.5)$, would exhibit a rather random
orientation of bond vectors, $P_2(\cos \theta) \approx 0$.

It is interesting to investigate this transition between a weakly compressed
semi-flexible polymer brush (where $\Psi = 0$), and the ``buckling deformation''
of individual stiff chains \cite{29,32,37,58}). One should bear in mind that
this transition depends on the various parameters of the problem: bending
stiffness $\kappa_b$, grafting density $\sigma_g$, and chain length $N$.

\begin{figure}[htb]
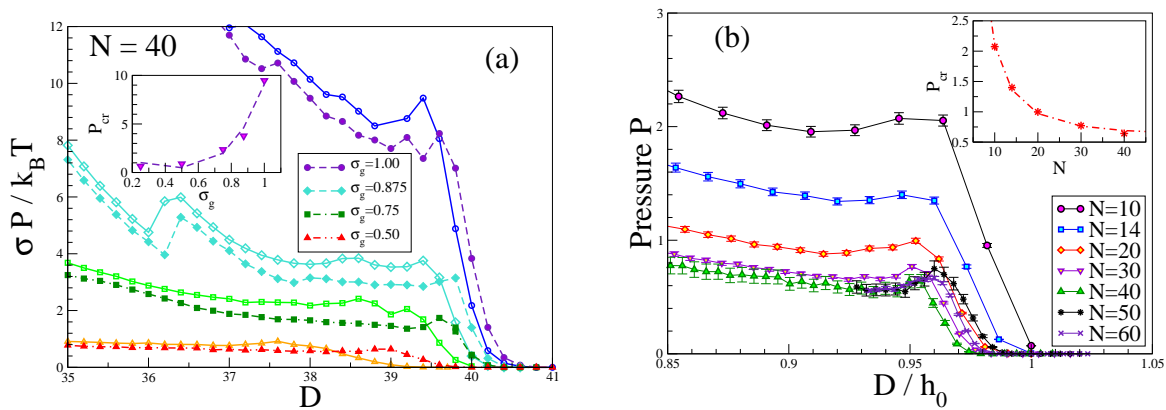

\vspace{0.7cm}
\includegraphics[scale = 0.30]{P_s_kb.eps}
\hspace{0.7cm}
\includegraphics[scale=0.30]{P_N.eps}

\caption{\label{fig10}
(a) Normal pressure $\sigma P/k_BT$ plotted vs. distance
$D$ for four choices of $\sigma _g$ and two choices of $\kappa_b$, $\kappa_b =
20$ (empty symbols) and $\kappa_b = 50$ (full symbols). All data are for $N=40$.
The inset shows the variation of the critical pressure $P_{cr}$  with
$\sigma_g$ for $\kappa_b = 20$. Dashed line denotes a fit $P_{cr} =
a_1 \sigma_g + a_2 \sigma_g^2 + a_3 \sigma_g^3$.
(b) Normal pressure $\sigma P/k_BT$ plotted vs. $D/h_0$ for the case $\kappa_b
= 50,\; \sigma _g = 0.5$, and a number of chain lengths from $N=10$ to
$N=60$, as indicated. The inset shows $P_{cr}$ vs chain length $N$, dashed line
is an empirical fit with $P_{cr} = c_1 N^{-2} + c_2$.}
\end{figure}

Fig.~\ref{fig10} presents a selection of our data on this issue. One can see 
that the onset of compression results in a steeper $\sigma P/k_BT$ curve, when 
$\sigma_g$ increases (Fig.~\ref{fig7}b), and also the almost flat plateau region
of $\sigma P/k_BT$, that is reached at compressions beyond the maximum, rises 
with $\sigma _g$ very distinctly. The variation with $\kappa_b$ is rather weak, 
however: the primary effect is a slight increase of the brush height $h_0$ with 
$\kappa_b$, while the height of the flat region in the pressure gets slightly 
smaller with increasing $\kappa_b$ (presumably the chains order the better the
stiffer they become). Moreover, the onset of bending occurs at {\em smaller}
degree of compression, the denser the brush, $\sigma_g$, is. An important detail
in this picture are also the observed occasional abrupt {\em slides} of the
brush chains, as in the $P - D$ curves for $\sigma_g = 0.875$ at $D \approx 36$,
whereby some excess elastic energy stored in the system is released. It is
conceivable that this effect is due to the finite (nonzero) rate of brush
compression in the course of the MD simulation.

A very pronounced effect is seen when the chain lengths is varied, however
(Fig.~\ref{fig10}b): Very stiff short chains exhibit much larger pressures for
comparable compressions $D/h_0$. Note that chains with $N=10$, $\kappa_b = 50$
are in the limit where the persistence length $\ell_p$ (which equals $\kappa_b$
for our choice of units) exceeds by far the contour length ${\cal L }\approx
0.96 (N-1)$, while for the chains with length $N=40$ to $60$ both lengths are
comparable. Interestingly, the data shown in Fig.~\ref{fig10}b, imply that the
pressure, when plotted versus $D/h_0$ for $D/h_0 <0.95$ (i.e., after the onset
of the $xy$-like ordering where chains bend uniformly in a chosen direction),
depends on the persistence length only, and not on the chain length (while $h_0$
does depend strongly on $N$, as shown already in Fig.~\ref{fig2}). The insets in
Fig.~\ref{fig10} present the variation of the critical pressure $P_{cr}$ with
grafting density. $P_{cr}$ itself is defined as the pressure $P$ where the
second-order transition, manifested by the onset of lateral orientation order
(in the $xy$-plane), takes place. For simplicity, we have estimated it roughly
from the pressure maximum (a more precise study of this transition, e.g., by
finite-size scaling methods \cite{56}, must be left to further work).

\begin{figure}[htb]
 \vspace{0.7cm}
 \includegraphics[scale=0.33]{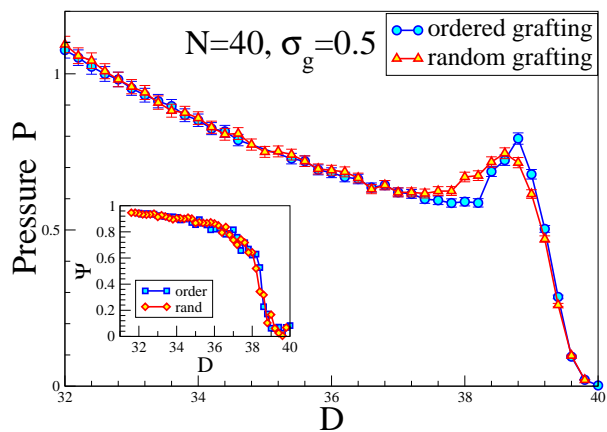}
 
 \caption{A comparison of polymer brushes with ordered (circles) and random 
(triangles) grafting in their response to external pressure. Here the 
brush comprises $512$ chains with $N=40$ and $\sigma_g=0.50$. The inset 
displays variation of the respective order parameter $\Psi$ with height $D$ of 
the compressing flat piston. \label{fig_ord_rand}}
\end{figure}

An important objection that could be raised against our findings is the 
suspicion that the regular square-lattice arrangement of our grafting sites 
leads to a structure of a dense semi-flexible brush that is much more regular 
than a real brush, for which the grafting sites are distributed at random. To
check this caveat, we also performed a simulation for a typical choice of
parameters ($N = 40,\; \sigma _g = 0.5, \; \kappa_b = 50$), where we compared
the pressure variation with compression for two brushes, one with an ordered
arrangement of grafting sites, and another one with an (almost) random grafting
- Fig.~\ref{fig_ord_rand}. By ``almost'' random, we mean that the randomly
chosen grafting sites were abandoned, if they were closer than a distance
$\sigma$ to a grafting site that was already present (so as to account for the
fact that the chemical groups that effect the grafting for two different chains
are subject to excluded-volume effective repulsion). 

One sees that the pressure variation with piston height $D$ for the random
choice of grafting sites hardly differs from  that for regular arrangement
except when the onset of order starts.  In this narrow interval the $P -
D$-relationship in the case of random grafting is slightly less sharp but rather
smeared over a slightly broader interval of heights $D$, presumably indicating
an effective ``dilution'' of the collective response due to randomness.

\section{A theoretical description}
\label{sec_Theory}

We start with the observation that for grafting densities $\sigma_g > 0.5$ and
stiff chains ($\kappa_b > 20$), the polymers in the uncompressed state (for
our choice of rather small chain length, $N < 100$) are stretched out to their
maximal length, that is, a rod-like conformation prevails. Transverse
fluctuations are very small, as the $xy$-components of the gyration radius
show - cf. Fig.~\ref{fig4}. Entropy plays then relatively little role and the
equilibrium properties of a compressed brush in this limit should follow from a
simple mechanical description.

\begin{figure}[htb]
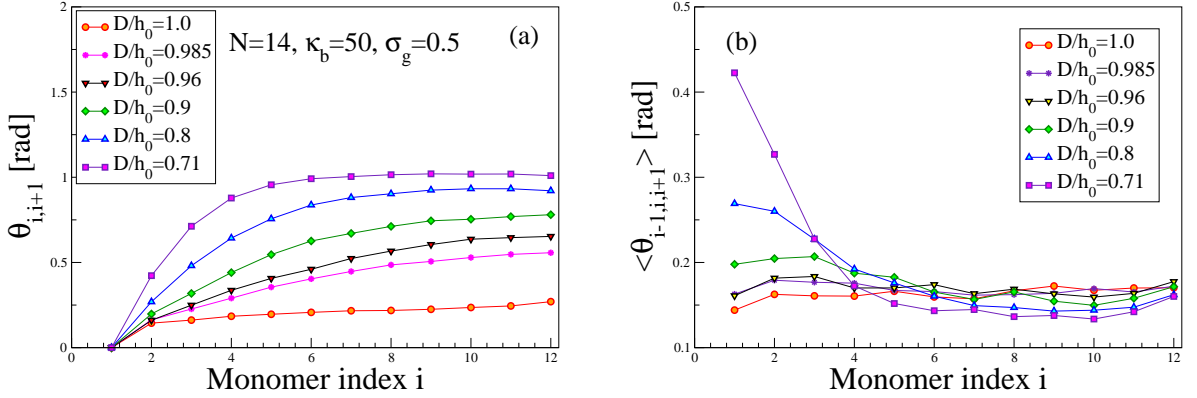

 \vspace{0.7cm}
\includegraphics[scale=0.30]{theta_i_N14_k_50_s0.5.eps}
\hspace{0.7cm}
\includegraphics[scale=0.30]{theta_neigh_N14_k_50_s0.5.agr.eps}

\caption{(a) Variation of the angle $\theta_{i,i+1}$ of consecutive bonds along
the chain backbone with the normal to the grafting surface at different
degrees of compression $D/h_0$, as indicated. Here $N=14,\; \sigma_g=0.5$, and
$\kappa_b = 50$. (b) The same as in (a) for the angle $\vartheta_{i-1,i,i+1}$
between neighboring bonds. \label{fig_angle} }
\end{figure}

The bending energy of a chain (including the uncompressed case) can be
approximated as
\begin{equation}
 U_{bend} = \sum_{i=1}^{N-1} \langle V_{b}(\vartheta_{i-1,i,i+1}) \rangle
\approx \frac{\kappa_b}{2} \sum_{i=1}^{N-1} \langle \vartheta_{i-1,i,i+1}^2
\rangle \approx \frac{\kappa_b}{2} N \langle \vartheta^2 \rangle
\label{eq_U_bend}
\end{equation}
In the last step one assumes that the distribution of bending energy along the
backbone of the chain is approximately uniform.  Fig.~\ref{fig_angle} supports
this conclusion, and also implies the very interesting result that $\langle
\vartheta_{i-1,i,i+1} \rangle \approx \langle \vartheta \rangle$ is independent
of the compression $D/h_0$, at least for small deformations $D/h_0 \ge 0.8$.
Recall from Fig.~\ref{fig10}b that for this choice of parameters the pressure
plateau (associated with the onset of lateral orientational order) starts at
$D/h_0 \approx 0.96$.

At zero pressure, and neglecting any interaction among the chains, one would
find (recall that for $P=0$ the azimuthal angle $\varphi$ is still random and
uniformly distributed, $\vartheta$ is a polar angle, and $\sin \vartheta \approx
\vartheta$ for $\kappa_b \gg 1$ can be used)
\begin{equation}
 \langle \vartheta \rangle \approx \int_0^\infty \vartheta ^2 \exp
\left (- \frac{\kappa_b}{2} \vartheta^2 \right)  d\vartheta / \int_0^\infty
\vartheta \exp \left (- \frac{\kappa_b}{2} \vartheta^2 \right ) d\vartheta
 = \sqrt{ \frac{\pi}{2\kappa_b} } 
\label{eq_theta}
\end{equation}
For the example shown in Fig.~\ref{fig_angle}b we would obtain $\langle
\vartheta \rangle \approx 0.17$, the value that is actually observed is close to
this value at $D/h_0 =1$. With increasing compression, larger angles for smaller
$i$ are observed, i.e., the rod-like chains undergo strong bend in the vicinity
of the substrate while staying straight away of this local deformation.

The opposite limit that we are considering is the case of perfect 
orientational ordering. Then the azimuthal angle $\varphi_{i,i+1}$ of all bonds
from monomer $i$ to monomer $i+1$ remains the same for all bonds of all chains, 
and the polar angles $\theta_{i,i+1}$ of the bonds with respect to the $z$-axis 
lie in a plane characterized by this azimuthal angle $\varphi$ and containing
the $z$-axis. In this limit a plausible assumption is that fluctuations of 
$\vartheta_{i-1,i,i+1}$ relative to their mean value $\langle \vartheta 
\rangle$ are small. Therefore we would conclude that
\begin{equation} \label{eq_U_mean}
 U_{bend} \approx \frac{\kappa_b}{2} N \langle \vartheta \rangle^2,\; \langle 
\theta_{i,i+1} \rangle = i \langle \vartheta \rangle, i=1,2,\ldots, N-1
\end{equation}
This result follows from the assumption that $U_{bend}$ has a minimum, subject
to the constraint that in the case of perfect azimuthal ordering the last angle
$\theta_{N-1,N} = \theta_{max}$ .Then one simply has to minimize the function
$\tilde{U} = U_{bend} - \lambda \sum_{i=1}^{N-1}\vartheta_{i-1,i,i+1}$ with
respect to all the $\vartheta_{i-1,i,i+1}$ with $\lambda$ being a Lagrange
multiplier. In this case one could derive a simple relation between $\langle
\vartheta \rangle$ and the distance $D$ of the topmost monomer from the surface,
as far as $\cos \langle \theta_{0,1} \rangle = 1$,  
\begin{equation} \label{eq_U}
 D = r_{min} \sum_{i=0}^{N-1} \cos \langle \theta_{i,i+1} \rangle 
\approx r_{min} \int_0^N \cos(s \langle \vartheta \rangle) ds =
r_{min} 
\frac{\sin(N\langle \vartheta \rangle)}{\langle \vartheta \rangle}
\end{equation}

Ignoring all other interactions among monomers, one then finds for the force 
$F$  exerted by  a chain in the brush on the compressing plane
\begin{equation}
 F = \frac{\partial U_{bend}}{\partial D} = \frac{\partial U_{bend}}{\partial 
\langle \vartheta \rangle} \frac{\partial \langle \vartheta \rangle}{\partial 
D} = \frac{\kappa_b}{r_{min}} 
\frac{\langle \vartheta \rangle^2}{\cos(N\langle \vartheta \rangle) - 
\frac{\sin(N\langle \vartheta \rangle)}{N\langle \vartheta \rangle} }
\end{equation}

The pressure $P$ would then be obtained simply by multiplying with the grafting 
density $\sigma_g$, to yield (for small $\langle \vartheta \rangle$) the result
\begin{equation}
 P = 3 \sigma_g \frac{\kappa_b}{r_{min}N^2}\left[ 1 + \frac{1}{10}  
\langle \vartheta \rangle ^2 + \ldots \right] \label{eq_P}
\end{equation}
while $1-D/h_0 \approx \frac{1}{6}\left( N \langle \vartheta \rangle \right)^2$
and $P_2(\cos\langle \theta \rangle) \approx 1 - \frac{1}{2} \left( N 
\langle \vartheta \rangle \right)^2$.

Our simulation data, however, imply that such a description (intended to hold 
for the limit $\kappa_b \to \infty$) still lacks an important ingredient which
we identify as the repulsive potential energy between the monomers. This energy
clearly increases when we compress the system (simply due to density increase).
Writing the volume fraction, occupied by the monomers, as $\phi = Nr_{min}^3
\sigma_g / D$, whereby $r_{min}^3$ is taken as an estimate of the volume of a
monomer, a mean-field estimate of the repulsive energy in the rod-like regime
that prevails for $P < P_{cr}$ yields:
\begin{equation}
 U_{rep} = v \phi^2 r_{min}^2 D = v \frac{N^2 \sigma_g^2 r_{min}^8}{D} = v
\frac{h_0^2 \sigma_g^2 r_{min}^6}{D} \label{eq_virial}
\end{equation}
where $v$ is a constant (proportional to the second virial coefficient).
The corresponding contribution to the pressure, $-\frac{\partial
U_{rep}}{r_{min}^2\partial D} = v \left (\frac{h_0}{D}\right )^2 \sigma_g^2
r_{min}^4$, has to be added to Eq.~(\ref{eq_P}) and then a revised prediction
for the critical pressure follows
\begin{equation}
 P_{cr} = \frac{3\sigma_g \kappa_b }{r_{min}} \frac{1}{N^2} + v \sigma_g^2 
r_{min}^4 \label{eq_Pcr}
\end{equation}

\begin{figure}[htb]
\vspace{0.7cm}
\includegraphics[scale=0.30, angle =270]{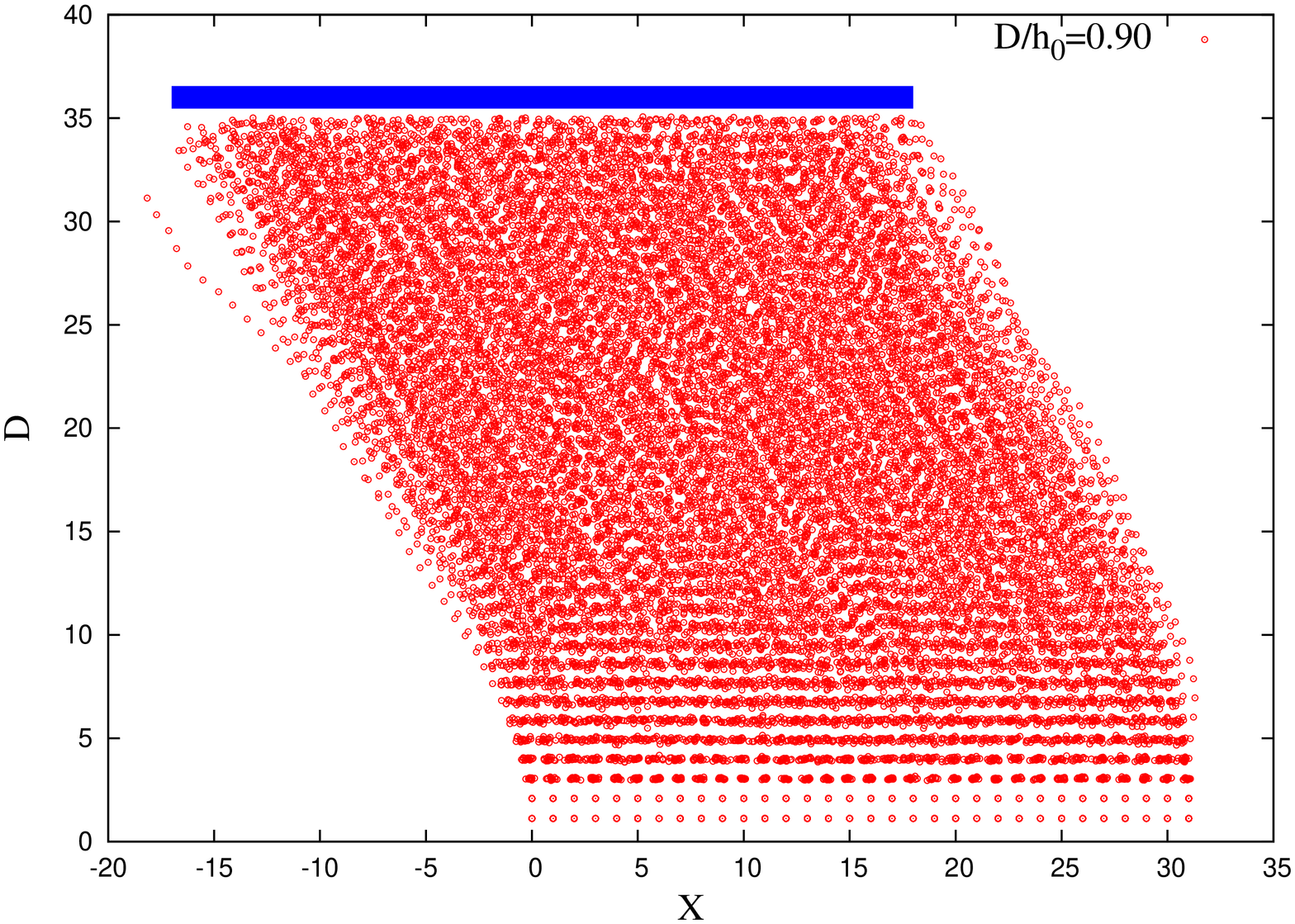}
\vspace{0.7cm}
\includegraphics[scale=0.30, angle =270]{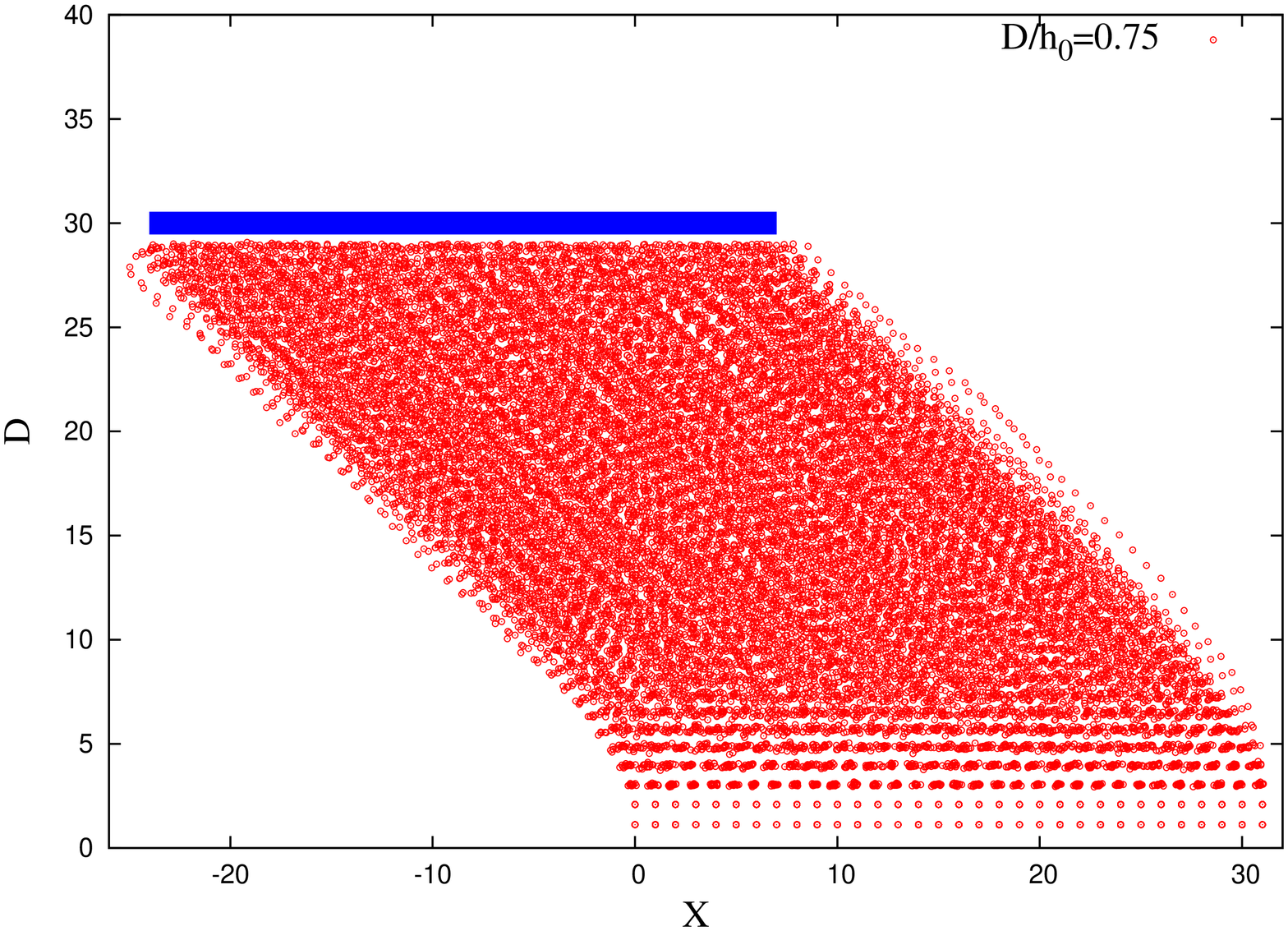}
 
\caption{Side view snapshots of a polymer brush with $N=40,\;\kappa_b=50$ at
$\sigma_g=0.50$ and two heights $D$ of the upper wall (indicated in blue).
(left) Degree of compression $D/h_0 = 0.90$, (right) $D/h_0 = 0.75$. Evidently,
with increasing compression the contour of the polymer chains in the  semi-rigid
brush develops a ``knickpoint'' close to the grafting surface, rather than a
regular arc that spans the two surfaces.  \label{fig_snap_def}}
\end{figure}

Fig.~\ref{fig_angle}a indicates, that the result $\langle \vartheta_{i-1,i,i+1}
\rangle = i\langle \vartheta \rangle $ holds only for very small degrees of
compression whereas for stronger compression,  $D/h_0 \le 0.9$, it does not
comply with our data. Snapshots of brush configurations  -
Fig.~\ref{fig_snap_def} - manifest instead that the polar angle $\theta_{i,i+1}$
attains its ultimate value $\theta_{max}$ in the narrow interval encompassing
the first few bonds only. Beyond this region of strong deformation in the
vicinity of the grafting surface, the compressed semi-rigid polymer brush
organizes itself into a quasi-crystalline nematic order of parallel bonds with
constant inclination angle $\theta_{max}$. 
 
Eq.~(\ref{eq_Pcr}) was the motivation for the fit, used in the inset of 
Fig.~\ref{fig10}b, in agreement with simulation measurements. In the inset of
Fig.~\ref{fig10}a, we found it useful to include a term that could be attributed
to the third virial coefficient and turn relevant in the limit of very large
grafting density $\sigma_g \to 1$.

\begin{figure}[htb]
\vspace{0.7cm}
 \includegraphics[scale=0.45, angle=0]{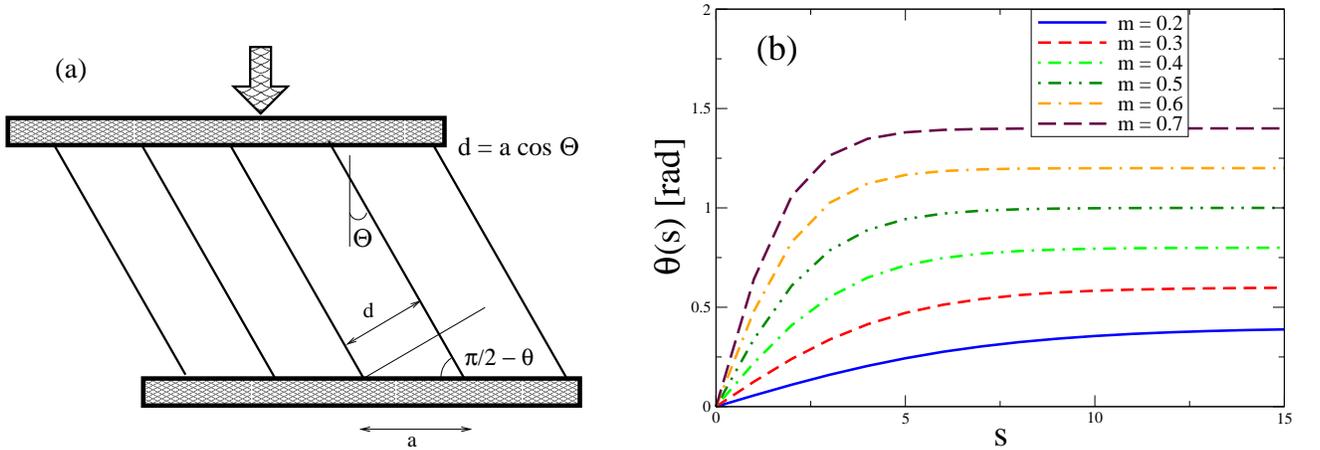}
 \hspace{0.7cm}
 \includegraphics[scale=0.35, angle=0]{theta_s.eps}
\vspace{0.5cm}

\caption{(a) As  the rod-like chains tilt at growing angle $\theta > 0$, the
distance $d$ between neighboring rods diminishes like $d = a \cos \theta$, where
$a$ is the separation between grafting points. (b) Variation of the polar angle
$\theta(s)=\frac{m}{w}\tanh \left ( \frac{m}{\sqrt{2}} s \right)$ along the
chain contour length of deformed polymer chains with $N = 14$ beads, $w=0.5$,
and different values of the parameter $m = \sqrt{\frac{P' - v'} {\kappa_b
\sigma_g}}$, as indicated. Apparently, the bending increases with growing
pressure $P'$, and even more so for softer, $\kappa_b \to 0$, or less dense
brushes. \label{fig_theta_s} }
\end{figure}

In what follows we replace the discrete index $i$ of each monomer by a
continuous variable: $\theta_{i,i+1} \to \theta(s)$.  From
Figs.~\ref{fig_angle}a,~\ref{fig_snap_def} we recognize that the bond
orientations $\theta(s)$ along the chain contour, cf. Eq.~\ref{eq_U}, describes
a nontrivial curve in the plane, singled out by the symmetry breaking of the
azimuthal angle $\varphi$ due to orientational order. If one would only take
the bending energy into account, one would obtain the trivial linear variation
$\theta(s) = s\langle \theta \rangle$ of Eq.~\ref{eq_U_mean}. In order to
derive a reasonable approximation for $\theta(s)$, it is necessary to take into
account both the energy due to compression, that is, a 'pressure times volume'
term, which is $U_{comp}(\theta) = P \sigma_g r_{min}^5 \int_0^N ds \cos
\theta(s)$, as well as the energy of monomer - monomer repulsion,
$U_{rep}(\theta)$. Indeed, with growing $P$, as the stiff chains bend more and
more, the monomers get closer to one another, unlike the case when the chains
are vertically stretched out like stiff rods. A simple geometric argument, cf.
Fig.~\ref{fig_theta_s}a, yields thus
\begin{equation} \label{eq_U_rep}
 U_{rep}(\theta) \equiv  v \sigma_g^2 r_{min}^7 \int_0^N
\frac{ds}{\cos
\theta(s)}
\end{equation}
For $\theta(s)=0$, Eq.~(\ref{eq_U_rep}) leads back to Eq.~(\ref{eq_virial}), of
course. Putting all terms together, the free energy that needs to be minimized,
becomes (using the abbreviations $P' = P\sigma_g r_{min}^5,\; v' = v\sigma_g^2
r_{min}^7$) becomes since the normal distance between rods in an array of
rods, tilted by angle $\theta$ scale like $\cos \theta^{-1}$ with respect to the
$z$-axis.
\begin{equation} \label{eq_F_NL}
 F = \int_0^N ds \left[ \frac{\kappa_b \sigma_g}{2} \left ( \frac{d\theta}{ds}
\right )^2 + \frac{v'}{\cos \theta(s)} + P' \cos \theta(s)  \right],
\end{equation}
which yields upon minimization with respect to $\theta (s)$ the following Euler
- Lagrange equation:
\begin{equation} \label{eq_EL}
 \kappa_b \sigma_g \frac{d^2\theta}{ds^2} - v' \frac{\sin \theta(s)}{\cos^2
\theta (s)} + P' \sin \theta(s) = 0.
\end{equation}

In the region where $\theta(s)$ is still small, one can expand Eq.~(\ref{eq_EL})
keeping only the terms up to order $\theta^3$, arriving at
\begin{equation}  \label{eq_KG}
 \frac{d^2\theta}{ds^2} + m^2 \theta(s) - w^2 \theta(s)^3 = 0,
\end{equation}
where the abbreviations
\begin{equation} \label{eq_m}
 m^2 \equiv \frac{P' - v'}{\kappa_b \sigma_g} ,\; w^2
\equiv \frac{P' + 5v'}{6 \kappa_b \sigma_g}
\end{equation}
have been introduced. Note that considerable deviations of $\theta$ from zero,
and the ensuing increased importance of the repulsion energy,
Eq.~(\ref{eq_U_rep}), emerge when the applied compressing pressure $P$ is 
sufficiently large so that the condition $P' > v'$, i.e., $m^2 > 0$  in
Eq.~(\ref{eq_m}) appears physically reasonable.

The approximation, involved in reducing Eq.~(\ref{eq_EL}) to Eq.~(\ref{eq_KG}),
holds for not too large angles $\theta$ only, and also Eq.~(\ref{eq_F_NL}) can
only hold at high grafting density $\sigma_g$ (otherwise the neglect of entropy
and the assumption of perfect symmetry breaking of the azimuthal angle cannot be
true). For sufficiently strong pressure, large angles $\theta > 1$ occur, and
one must then solve the full Eq.~(\ref{eq_EL}).

The governing Eq.~(\ref{eq_KG}), derived in the present treatment, is familiar
from the so called $\phi^4$-model of statistical mechanics, and has emerged in
different contexts. With respect to the boundary condition, $\theta(s \to 0) =
0$, one obtains the well-known soliton (kink) solution
\begin{equation} \label{eq_kink}
 \theta(s) = \frac{m}{w} \tanh \left (\frac{m}{\sqrt{2}} s \right) =
 \sqrt{\frac{P'-v'}{P'+5v'}} \tanh \left ( \sqrt{\frac{P'-v'}{2\kappa_b \sigma_g}} s \right ).
 \end{equation}
where $m$ and $w$ are positive constants, cf.  Eq.~(\ref{eq_m}). Indeed,
Fig.~\ref{fig_theta_s} demonstrates that the solution, Eq.~(\ref{eq_kink}),
nicely reproduces the observed MD data, shown in Fig.~(\ref{fig_angle})a, at
least qualitatively.  Eventually, one should note that going back to the case of
small deformation when $\theta \to 0$, and the energy contribution due to chain
repulsion, Eq.~(\ref{eq_U_rep}), can be neglected, a solution of our governing
equation, Eq.~(\ref{eq_EL}), recovers the expression for the critical pressure
$P_{cr}$, namely, Eq.~(\ref{eq_Pcr}), as expected \cite{24}. It is worth
mentioning, that this result, Eq.~(\ref{eq_Pcr}), has been first obtain by L.
Euler and D. Bernoulli for the famous 'beam bending' problem in the year $1750$.

\section{Outlook on dynamic behavior}
\label{sec_Dynamics}

In polymer brushes one may study relaxation phenomena on many different levels:
(i) fluctuations of individual monomers and and motion of chains as a whole in
thermal equilibrium \cite{Reith}, (ii) collective relaxation phenomena as
described by dynamic correlations of the local monomer density in the brush,
(iii) the nonlinear dynamic response, associated with compression and shear,
e.g. \cite{10}. A comprehensive treatment for any of these topics for brushes
formed from rather stiff chain molecules is beyond the scope of the present
paper 
\begin{figure}[htb]
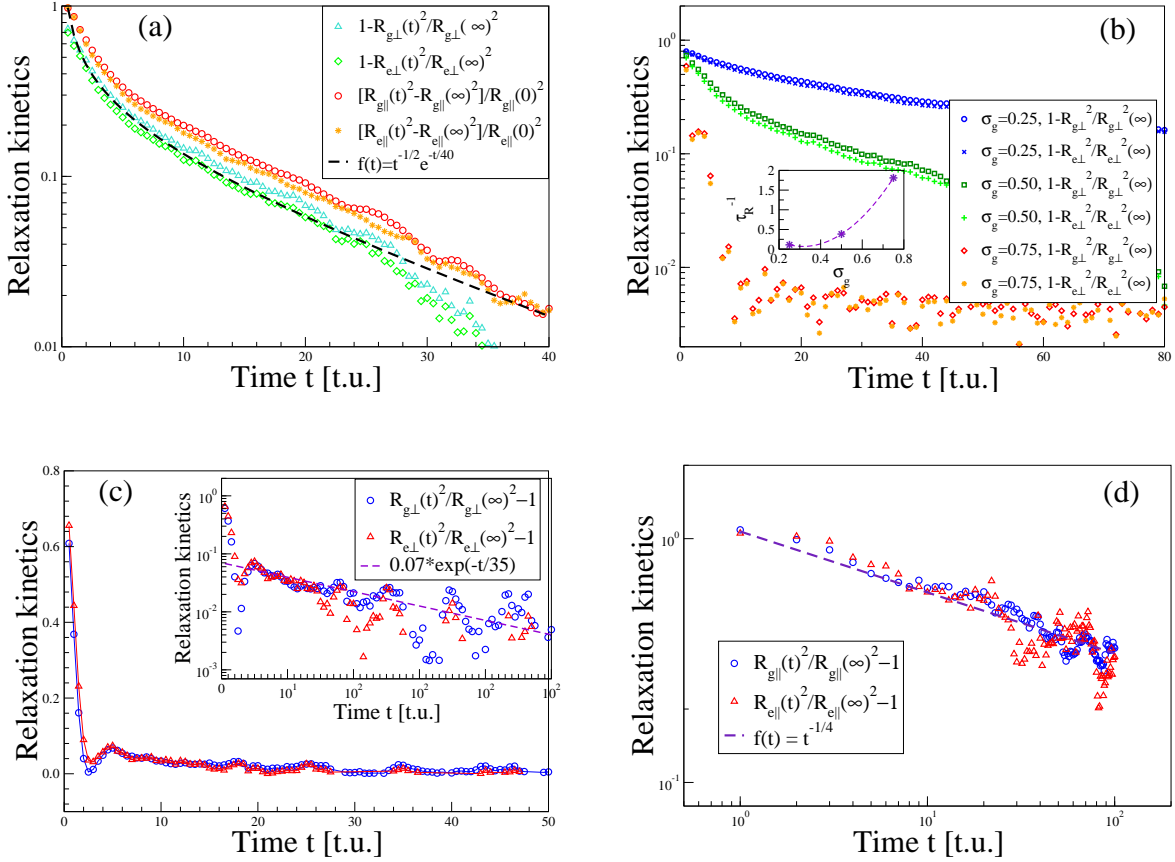

\vspace{0.7cm}
\includegraphics[scale=0.30]{kin_s_0.5_stiff.eps}
\hspace{0.7cm}
\includegraphics[scale=0.30]{kin_rg_re_N40.eps}
\vspace{0.9cm}

\includegraphics[scale=0.30]{kin_s_0.5_flex_perp.eps}
\hspace{0.7cm}
\includegraphics[scale=0.30]{kin_s_0.5_flex_par.eps}

\caption{\label{fig_kin} Relaxation functions of perpendicular and parallel mean
square components of the gyration radius of the chains. The polymer brush
recovery was monitored after the compressing piston was instantaneously removed
at time $t=0$. All cases shown refer to $N=40$, $\kappa_b=20$, and $D(t=0)/h_0=
1/2$. Case (a) shows $\sigma _g =0.5$, while (b) compares three grafting
densities, as indicated. The characteristic relaxation time $\tau_R$ for
initial brush recovery is shown in the inset as function of $\sigma_g$ along
with a fit by $\tau_R^{-1} = b_0+b_1 \sigma_g + b_2 \sigma_g^2$. A comparison
with the recovery kinetics of a polymer brush at $\sigma_g=0.5$, comprised of
totally flexible chains, $\kappa_b=0$, is presented in (c) and (d).
Here again $N=40$.}
\end{figure}
yet here we present few results which relate to the relaxation (brush recovery),
associated with changes of the compression $D/h_0$ of the brush (recall the
hysteresis observed near the onset of lateral orientational ordering). A very
straightforward computer experiment, which could also possibly carried out in
the laboratory, considers the relaxation of a compressed brush after the
compressing piston at time $t = 0$ is suddenly removed and the brush relaxes
from its height $h(t=0)=D$ back to its uncompressed state with height $h_0$.
Fig.~\ref{fig_kin} shows the time dependence of the $z$- and $xy-$components of
both the end-to-end vector and gyration radius square for few typical cases of
such computer experiment. The data are shown as semi-logarithmic plots which
would lead to straight lines, if the relaxation functions plotted were simple
exponential (which obviously is not the case). Fig.~\ref{fig_kin}a indicates
that the relaxation behavior of the $xy-$ and $z$-components is similar, and
Fig.~\ref{fig_kin}b shows that the initial decay becomes much faster with
increasing grafting density. Fig.~\ref{fig_kin} includes also an empirical fit
to  a function $const. t^{-1/2} \exp(-t/\tau_R)$ where the relaxation time
$\tau_R$  is a second adjustable parameter, $\tau_R \approx 40$.

These measurements are compared to those concerning time recovery of an
initially compressed brush comprised of flexible chains, $\kappa_b = 0$, and
otherwise identical parameters -  Fig.~\ref{fig_kin}c,~d. Apparently, in the
absence of chain stiffness the relaxation kinetics for the  $z$ and
$xy-$components differs significantly: components perpendicular to the grafting
surface undergo short-termed fast recovery followed by a much slower secondary
relaxation whereas the relaxation of the parallel components appears to be
governed by a power law $\propto t^{-1/4}$ -   Fig.~\ref{fig_kin}d.


\section{Conclusions}
\label{sec_Summary}

In this work Molecular Dynamics simulations of dense polymer brushes, formed
from semi-flexible chains were presented, focusing on the regime where
persistence length and contour length of these rather stiff macromolecules are
comparable. In addition, the chemical nature of the grafted first bond renders
it perpendicular to the underlying surface and, therefore, shapes the overall
brush behavior.

We show that this behavior is very different from what is known
for brushes comprised of flexible chains. It turns out that the chains in the
brush at high grafting density behave nearly as rigid rods, and the
$z$-components of the chain end-to-end distance and gyration radius square take
their maximal possible values, provided the chemistry of chain grafting fixes
the first bond perpendicular to the substrate. 

It is shown that fluctuations of the chain in the transverse $(xy)$-directions,
perpendicular to the $z$-axis along which the rods are oriented, are extremely
small for large bending stiffness, and decrease like a simple exponential with
growing grafting density $\sigma_g$ - cf. Fig.~\ref{fig4}a. In contrast, the
lateral size of a chains, $R_{gxy}^2, \; R_{xy}^2$, changes non-monotonically
with stiffness and goes through a maximum for a finite degree of rigidity. This
behavior is very different from what is known for flexible chains in brushes in
the case of semi-dilute and concentrated monomer densities in the brush. Only
for the case of melt densities, crystallization in polymer brushes has been
predicted \cite{Merlitz} which may be related to our observations.

In the present work we have focused on the compression of these stiff chain
brushes and on the resulting orientational order of the bonds in the
$xy$-directions during bending. We have demonstrated that the onset of this
order is a continuous phase transition where the rotational symmetry in the
$xy$-plane is broken. However, since all bonds of the chains in a perfectly
ordered system would have to share this ordering in the simulation box,
requiring large lateral monomer displacements in the same direction, there is
obviously a commensurability conflict between the lateral deformation of such a
brush and the periodic boundary conditions. This singles out the $x,\; y$ axes,
imposing a cubic symmetry of the bending phenomenon. Due to these problems, a
study of the critical behavior of this phase transition would be premature, and
has not been attempted.

In order to obtain theoretical guidance on the observed phenomenon, some
phenomenological considerations were presented. Again we emphasize that these
considerations and the resulting interesting predictions, cf. Section
\ref{sec_Theory}, should be taken as a first step only while a more systematic
theory even on the mean-filed level remains a challenge for the future.

One interesting feature is the finding that for large compression the resilient
pressure of such stiff chain brushes, exerted on the compressing piston, is
much smaller that for flexible chain brushes of the same chain length and
grafting density. Thus, a study of shear forces between two stiff chain
brushes would also be very interesting but must be left as a future task too.

In conclusion, we hope that the present work will stimulate experimental
interest in this subject. We also believe that some related phenomena may take
place in dense arrangements of stiff biopolymers in the context of complex
biological structures.

\section{Acknowledgments}
We are indebted to Friederike Schmid, Oleg Borisov, and Burkhard D\"{u}nweg for
instructive discussions of this work. This investigation was supported by the
Deutsche Forschungsgemeinschaft (DFG) under Grant No Bi 314/23.

\clearpage

For Table of Contents Use Only
\vspace*{0.50cm}

{\bf Unconventional ordering behavior of semi-flexible polymers in dense
brushes under compression}
\vspace{1.0cm}
 by Andrey Milchev, and Kurt Binder

\vspace{1.0cm}

\includegraphics[scale=0.45]{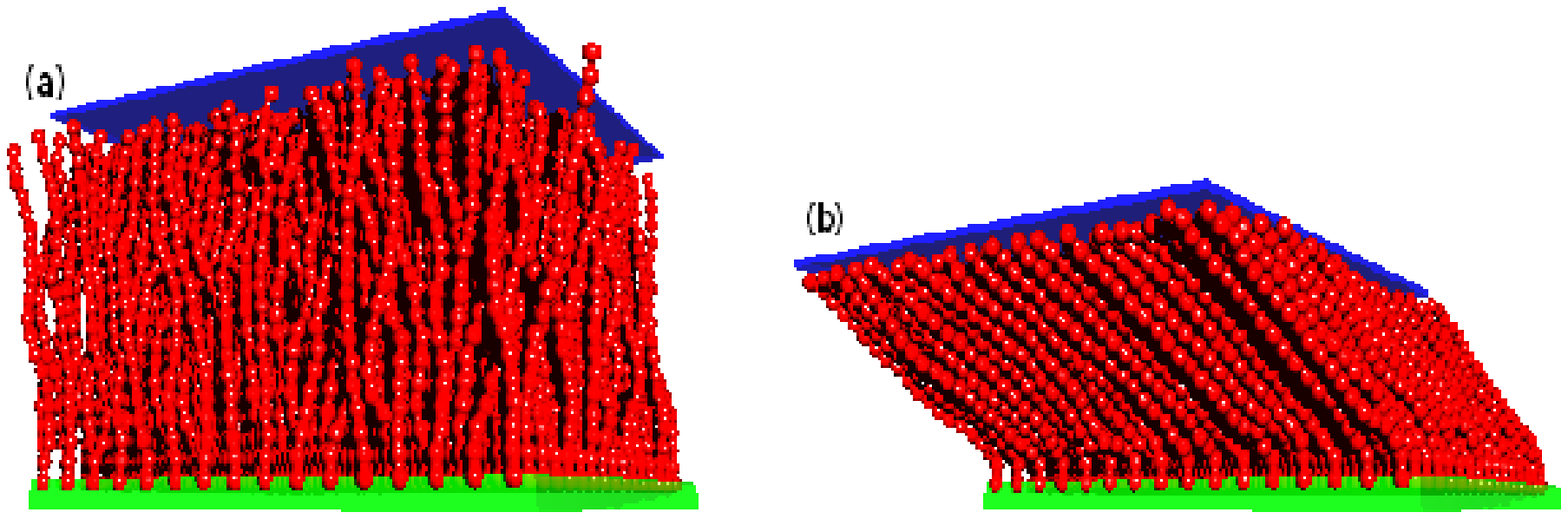}
\vspace{1.0cm}

TOC graphics: Snapshots of a compressed semi-flexible brush under flat piston 
at two different degrees of deformation. 
\end{document}